\pdfoutput=1
\RequirePackage{ifpdf}
\ifpdf 
\documentclass[pdftex]{sigma}
\else
\documentclass{sigma}
\fi

\def\w{\omega}
\def\g{\gamma}
\def\S{\Sigma}
\def\s{\sigma}
\def\t{\tau}
\def\a{\alpha}
\def\b{\beta}
\def\m{\mu}
\def\n{\nu}
\def\D{\Delta}
\def\d{\delta}
\def\r{\rho}
\def\l{\lambda}
\def\th{\theta}

\def\e{\epsilon}

\def\p{\phi}
\def\he{\hat =}

\numberwithin{equation}{section}

\begin{document}

\allowdisplaybreaks

\renewcommand{\thefootnote}{$\star$}

\renewcommand{\PaperNumber}{005}

\FirstPageHeading

\ShortArticleName{Entropy of Quantum Black Holes}

\ArticleName{Entropy of Quantum Black Holes\footnote{This
paper is a contribution to the Special Issue ``Loop Quantum Gravity and Cosmology''. The full collection is available at \href{http://www.emis.de/journals/SIGMA/LQGC.html}{http://www.emis.de/journals/SIGMA/LQGC.html}}}

\Author{Romesh K.~KAUL}

\AuthorNameForHeading{R.K.~Kaul}

\Address{The Institute of Mathematical Sciences, CIT Campus, Chennai-600 113, India}
\Email{\href{mailto:kaul@imsc.res.in }{kaul@imsc.res.in}}

\ArticleDates{Received September 14, 2011, in f\/inal form February 03, 2012; Published online February 08, 2012}

\Abstract{In the Loop Quantum Gravity, black  holes (or even more general Isolated Horizons)
are described by a  $SU(2)$ Chern--Simons theory.  There is an equivalent formulation of
the horizon degrees of freedom in terms of a $U(1)$ gauge theory which is just
a gauged f\/ixed version of the $SU(2)$ theory.  These developments
will be surveyed here. Quantum theory based on either formulation can be used to count the
horizon micro-states associated with quantum geometry f\/luctuations and
from this the   micro-canonical entropy can be obtained. We shall review the computation in
$SU(2)$ formulation.  Leading term in the entropy is proportional to
horizon area with a coef\/f\/icient depending  on the Barbero--Immirzi parameter
which is f\/ixed by matching  this result with  the Bekenstein--Hawking formula.
Remarkably there are corrections beyond the area term,
the leading one is logarithm of the horizon area with a def\/inite coef\/f\/icient $-3/2$, a result
which is more than a decade old now. How the same results are obtained in the equivalent
$U(1)$ framework  will also be indicated. Over years, this entropy formula  has also been
arrived at from a variety of other perspectives. In particular, entropy of BTZ black holes
in three dimensional gravity exhibits the same logarithmic correction.
Even in the String Theory, many  black hole models   are known to possess such properties.
This suggests a possible universal  nature of  this logarithmic correction.}


\Classification{81T13; 81T45; 83C57; 83C45; 83C47}

\renewcommand{\thefootnote}{\arabic{footnote}}
\setcounter{footnote}{0}

\section{Introduction \label{I}}
Black holes have fascinated the imagination of physicists and astronomers for a long time now.
 There is mounting  astronomical evidence for  objects with black hole like properties; in fact,
these  may occur  abundantly   in  the Universe. Theoretical studies of
black hole properties have been pursued, both at the classical level and traditionally at
semi-classical level,  for a long time. The pioneering work of Bekenstein, Hawking
and others during seventies of the last century have suggested that black holes are endowed
with thermodynamic attributes such as entropy and temperature \cite{bh}.
Semi-classical arguments have led to the fact that  this entropy is very large  and is given,
in the natural units, by a quarter of  the horizon area, the Bekenstein--Hawking
area law.  Understanding  these properties is a fundamental challenge
 within the framework of a full f\/ledged theory of quantum gravity. The entropy would
 have  its origin in the quantum gravitational  micro-states associated with
 the horizon. In fact reproducing  these thermodynamic properties
 of black holes  can be considered as a possible test of
 such a quantum theory.

 There are  several proposals for theory of quantum gravity. Two of these are the String
 Theory and the  Loop Quantum Gravity. There are other theories like dynamical triangulations
  and also Sorkins's causal set framework.
  Here we shall survey some of the  developments
 regarding black hole entropy  within a particular
 theory of quantum gravity, the Loop Quantum Gravity (LQG) where
 the degrees of freedom of the event horizon of a black hole are described by a quantum
$SU(2)$ Chern--Simons theory.
 This also holds for the more general horizons, the Isolated Horizons  of Ashtekar et al.~\cite{ash1},
 which, def\/ined quasi-locally,  have been introduced to describe situations like a black hole  in equilibrium with
 its  dynamical exterior. Not only is the semi-classical
 Bekenstein--Hawking area law reproduced for a large   hole, quantum micro-canonical entropy
 has additional corrections which depend on
 the logarithm of horizon area with a def\/inite,  possibly universal,  coef\/f\/icient $-3/2$,
 followed by an area independent constant and terms which are inverse powers of area.
Presence of these additional
 corrections is the hallmark of   quantum geometry.  These
results,  f\/irst derived within  LQG  framework in four dimensions,
 have also   been seen to emerge in other  contexts. For
example,  the entropy of   BTZ black holes in three-dimensional gravity   displays similar properties.
Additionally, application of the  Cardy formula of conformal f\/ield theories, which are relevant
to study   black holes in the  String Theory,  also implies such corrections to the area law.
Though the main thrust of this  article  is to survey developments in LQG, we shall
 also review, though only brief\/ly, a few calculations of black hole entropy from other perspectives.

 \section{Horizon topological f\/ield theory \label{II}}

 That the horizon degrees of freedom  of a black hole are   described by a $SU(2)$
 topological f\/ield  theory follows readily from following two facts \cite{rkpm}:

 (i) The event horizon (EH) of a black hole space-time  (and more generally an Isolated
Horizon (IH)~\cite{ash1}), is a null inner boundary of the
  space-time   accessible to an asymptotic observer. It   has
 the topology $\mathbb{R} \times \mathbb{S}^2$ and {\it a degenerate intrinsic three-metric}.
 Consequently, such a manifold
can not support any local propagating degree of freedom which would, otherwise,
have to be described by a Lagrangian density containing determinant and
inverse of the metric. {\it The horizon degrees of freedom have to be entirely
global or topological. These  can be  described only by a  theory which does not depend
on the metric, a topological  quantum field theory}\footnote{For   reviews
of topological f\/ield theories see, for example~\cite{th, rk}.}.

(ii) In the Loop Quantum Gravity   framework, bulk space-time properties
are described in terms of Sen--Ashtekar--Barbero--Immirzi
  {\it real} $SU(2)$ connections~\cite{rovbk}. Physics associated with bulk
space-time geometry is invariant under local $SU(2)$ transformations. The
EH (more ge\-ne\-rally the IH) is   a  null  boundary   where Einstein's equation
  holds.
{\it At the classical level, the degrees of
freedom and their dynamics on an EH $($IH$)$ are   completely determined by the
  geometry and dynamics in the bulk. Quantum theory of  horizon degrees of freedom
has to imbibe this $SU(2)$ gauge invariance from the bulk.}

  In view of these two properties,  degrees of freedom associated with a   horizon
     have to be described by a topological f\/ield theory
  exhibiting $SU(2)$ gauge
  invariance. There are two such three-dimensional candidates, the Chern--Simons
  and BF theories. However, both these theories essentially capture the
  same topological properties~\cite{rk}
   and hence would  provide equivalent descriptions.
 It is, therefore, no surprise  that when the detail properties of the various
 geometric quantities on the horizon are analysed, as has been done in several
places in literature,   they are found to obey  equations of motion
 of  the topological $SU(2)$ Chern--Simons theory (or equivalently BF theory)
 with specif\/ic sources on the three-manifold
$\mathbb{R}\times \mathbb{S}^2$. This description can be presented either in the form of
a theory with full f\/ledged $SU(2)$ gauge invariance or, equivalently, by a
gauge f\/ixed $U(1)$ theory. We shall review this   in Section~\ref{II.1} below for the
Schwarzschild hole. Similar results hold for the more general case of
Isolated Horizons~\cite{ash1}, which shall be brief\/ly summarized   next in the
Section~\ref{II.2}. The sources of the Chern--Simons  theory     are  constructed
from tetrad components in the bulk. Clearly,  quantizing this Chern--Simons theory
paves a way  for counting    micro-states of  the  horizon and hence the
associated entropy which we shall take up in   Section~\ref{III}.

\subsection{Schwarzschild black hole} \label{II.1}

Following closely the analysis of~\cite{rkpm}, we shall study the
properties of  future event horizon of the Kruskal--Szekeres extension of
Schwarzschild space-time explicitly. In the process, we shall unravel the relationship
between the horizon $SU(2)$ and $U(1)$
Chern--Simons theories.
We display an appropriate set of   tetrad f\/ields, which f\/inally leads to the gauge f\/ields
on the black hole (future) horizon with only manifest $U(1)$ invariance. For this choice,
  we f\/ind that
  two components of $SU(2)$ triplet solder forms on the spatial slice of the horizon,
orthogonal to the direction specif\/ied by the $U(1)$ subgroup,
   are indeed
zero as they should be. This is in agreement with the    general analysis of~\cite{bkm}.
In the next subsection, we explicitly demonstrate how the equations of motion of   $U(1)$ theory so
obtained are to be interpreted as those coming from a     $SU(2)$ Chern--Simons theory
through a partial gauge f\/ixing procedure.
In the course of our analysis, we also derive the dependence of coupling
constant of these Chern--Simons  theories on the Barbero--Immirzi parameter $\g$
and  horizon area $A_{\rm H}$.

 Schwarzschild metric  in  the Kruskal--Szekeres null coordinates $v$ and $w$  is
 given by its non-zero components as:
$g^{}_{vw}  =   g^{}_{wv}= - (4 r^3_0 / r ) \exp \left( -   r/ r^{}_0 \right)$,
$g^{}_{\th\th} = r^2_{}$, $g^{}_{\p \p} = r^2_{} \sin \th$.
 Here $r$ is a~function of~$v$ and~$w$ through:
$- 2 v w=  \left[(   r /r^{}_0)- 1\right] \exp \left(   r /r^{}_0 \right)$.
  An appropriate set of tetrad f\/ields  compatible with this
metric  in the exterior region  of the black hole($v>0$, $w<0)$ is:
\begin{alignat}{3}
& e^0_\m  =
  {\sqrt {\frac A  2}}\left( {\frac w {\a}}  \partial_\m^{}v +{\frac {\a} w}  \partial^{}_\m w \right),\qquad &&
  e^1_\m  =     {\sqrt {\frac A  2}}\left({\frac  w {\a}}  \partial_\m^{}v -{\frac {\a}w}
 \partial^{}_\m w \right), & \nonumber \\
& e^2_\m  =    r  \partial^{}_\m \th , \qquad &&
e^3_\m  =   r\sin \th  \partial^{}_\m \p, & \label{tetradsexKruskal}
\end{alignat}
where $A \equiv (4 r^3_0/ r) \exp \left( -    r  /r^{}_0 \right)$ and $\a$ is an arbitrary
 function of the coordinates. A choice of~$\a(x)$
characterizes the local Lorentz frame in the indef\/inite metric plane ${\cal I}$ of the Schwarzschild space-time
whose spherical symmetry implies that it has the topology ${\cal I} \otimes \mathbb{S}^2$.
  The spin connections   can be constructed for this set of tetrads to be:
\begin{gather}
 \w^{01}_\m  =  -{\frac 1 2} \left( 1-{\frac {r^2_0} {r^2_{}}}\right) {\frac 1 v}   \partial^{}_\m v -
{\frac 1 2} \left( 1+ {\frac {r^2_0} {r^2_{}}} \right) {\frac 1 w} \partial_\m w + \partial_\m \ln \a,
\qquad  \w^{23}_\m = -\cos \theta  \partial^{}_\m \p,  \nonumber \\
 \w^{02}_\m = - {\sqrt {\frac A 2}}  {\frac 1 {2r^{}_0}} \left( {\frac
     {vw} {\a}}  + \a \right) \partial_\m \th ,  \qquad \w^{03}_\m =  -
 {\sqrt {\frac A 2}}  {\frac {\sin \th} {2r^{}_0}} \left( {\frac {vw} {\a}} +
  \a \right) \partial_\m \p ,  \nonumber \\
\w^{12}_\m =   - {\sqrt {\frac A 2}}  {\frac 1 {2r^{}_0}}   \left( {\frac
    {vw} {\a}} - \a \right) \partial_\m \th ,  \qquad
\w^{13}_\m=   - {\sqrt {\frac A 2}}  {\frac {\sin \th} {2r^{}_0}} \left(
  {\frac {vw} {\a}} - \a \right) \partial_\m \p .  \label{spinconnKruskal}
\end{gather}

   LQG is described in terms of  linear combinations of these connection components in\-vol\-ving
the Barbero--Immirzi parameter
 $\g$, the real Sen--Ashtekar--Barbero--Immirzi  $SU(2)$ gauge   f\/ields~\cite{rovbk}.
   To this ef\/fect, we construct the  $SU(2)$ gauge f\/ields:
 \begin{gather}
 A^{ (i)}_\m =    \g \w^{0i}_\m  -  {\frac 1 2} \e^{ijk}_{} \w^{jk}_\m . \label{gaugeA}
\end{gather}

The black hole horizon is the future horizon   given by $w  =0$. This
 is a null three-manifold $\D$ which is topologically $\mathbb{R}\times \mathbb{S}^2_{}$ and is described
by the coordinates $a= (v , \theta, \p)$ with $ 0 <  v < \infty$, $0\le
\theta < \pi$, $0\le \p<2\pi$. The foliation of   manifold $\D$ is provided by
$v=constant$ surfaces,   each an~$\mathbb{S}^2$.
The relevant tetrad f\/ields~$e^I_a$   on  the horizon $\D$
from~(\ref{tetradsexKruskal}) are:
$e^0_a  \,\he\, 0$, $e^1_a  \,\he\,   0$, $ e^2_a  \,\he\,  r^{}_0  \partial^{}_a \th$,
 $  e^3_a  \,\he\,  r^{}_0 \sin \th   \partial^{}_a \p$ where $a= (v, \th,\p)$
(we denote equalities on $\D$, that is for $w=0$,
 by the symbol~$\he$). The intrinsic metric on $\D$  is degenerate with its signature
$(0,+,+)$   and is given by $q^{}_{ab} = e^I_a e^{}_{Ib}
\,\he\, m^{}_a {\bar m}_b + m^{}_b {\bar m}^{}_a$  where $m^{}_a \equiv
  r^{}_0  \left( \partial^{}_a \th+ i \sin \th  \partial^{}_a \p \right)/\sqrt 2 $.

Only non-zero solder form $\S^{IJ}_{ab}\equiv e^I_{[a} e^J_{b]} $
on the horizon is $ \S^{23}_{ab} \,\he\, r^2_0 \sin \th \partial^{}_{[a} \th \partial^{}_{b]} \p$.
 The spin connection f\/ields from (\ref{spinconnKruskal}) are:
\begin{alignat}{4}
& \w^{01}_a \, \he\,  \frac{1}{2}  \partial^{}_a \ln \b,\qquad && \w^{02}_a \, \he\,  - \sqrt{\b}  \partial^{}_a \th,
\qquad &&   \w^{03}_a \, \he\,  -\sqrt{\b} \sin \th \partial^{}_a  \p, & \nonumber \\
& \w^{23}_a \, \he\,  - \cos \th  \partial^{}_a \p ,\qquad && \w^{31}_a \, \he\,
- \sqrt{\b} \sin \th  \partial^{}_a \p  , \qquad &&  \w^{12}_a \, \he\,  \sqrt{\b} \partial^{}_a \th, & \label{connection}
\end{alignat}
where $\b =  \a^2_{}/(2e)$ with $e \equiv \exp(1)$. Notice that the spin connection f\/ield $\w^{01}_a
=\frac{1}{2} \partial^{}_a \ln \b$
here, with a possible singular behaviour for $\b=0$, is a pure gauge. If we wish,
by a suitable boost transformation $\w^{IJ}_a \rightarrow \w'^{IJ}_a$,
it can be rotated away to zero, with corresponding changes in other spin connection f\/ields:
$\w'^{01}_a = \w^{01}_a - \partial^{}_a \xi$, $ \w'^{23}_a = \w^{23}_a$, $\w'^{02}_a
= \cosh \xi \w^{02}_a + \sinh \xi \w^{12}_a$,
$\w'^{03}_a = \cosh \xi \w^{03}_a + \sinh \xi \w^{13}_a$,
$ \w'^{12}_a = \sinh \xi \w^{02}_a+ \cosh \xi  \w^{12}_a$,
and $ \w'^{13}_a   = \sinh \xi \w^{03}_a + \cosh \xi  \w^{13}_a$. For the choice
$\xi = \frac{1}{2} \ln \left( \b /\b'\right)$ with $\b'$ as an arbitrary constant, this leads to
$\w'^{01}_a \, \he\, 0$, $\w'^{23}_a \, \he\, - \cos \th \partial^{}_a \p$,
$ \w'^{02}_a \, \he\,  - \sqrt{\b'} \partial^{}_a \th$, $\w'^{03}_a \,\he\,  -\sqrt{\b'}
\sin \th \partial^{}_a  \p$, $\w'^{12}_a \,\he\,  \sqrt{\b'} \partial^{}_a \th $ and
$\w'^{13}_a \,\he\,  \sqrt{\b'} \sin \th \partial^{}_a \p$.

To demonstrate that the horizon degrees of freedom can
be described by a Chern--Simons theory, we  use (\ref{connection}) to write
the relevant components of   $SU(2)$ gauge f\/ields (\ref{gaugeA}) on $\D$ as:
 \begin{gather}
 A^{(1)}_a \, \he\,  {\frac {\g} 2}  \partial_a \ln \b+\cos \th \partial_a \p, \qquad
 A^{(2)}_a \, \he\,  -    {\sqrt {\b}}
  \left( \g  \partial_a \th  -  \sin \th  \partial_a \p \right), \nonumber \\
  A^{(3)}_a \, \he\,  -   {\sqrt {\b}}
  \left( \g \sin \th   \partial_a \p + \partial_a \th \right). \label{A'}
\end{gather}
    The    f\/ield strength
components constructed from these  satisfy the following relations
  on $\D$:
\begin{gather}
  F^{(1)}_{ab} \equiv 2\partial^{}_{[a }A^{(1)}_{b]}
  + 2A^{(2)}_{[a} A^{(3)}_{b]}   \, \he\, - {\frac 2 {r^2_0}}  \left( 1 -  K^2_{} \right)  \S^{23}_{ab},
\nonumber \\
 F^{(2)}_{ab} \equiv 2\partial^{}_{[a} A^{(2)}_{b]}
  +2 A^{(3)}_{[a} A^{(1)}_{b]}  \,\he\,  - 2 {\sqrt {1+\g^2_{}}} \sin \th \partial_{[a}
\p \partial_{b]} K, \nonumber \\
F^{(3)}_{ab} \equiv 2 \partial^{}_{[a } A^{(3)}_{b]}
  + 2A^{(1)}_{[a} A^{(2)}_{b]}  \,\he\,  2{\sqrt {1+\g^2_{}}} \partial^{}_{[a} \th \partial_{b]} K,
  \label{F'}
\end{gather}
where  $\S^{IJ}_{\m\n} = e^I_{[\m} e^J_{\n]} \equiv \frac{1}{2}\left( e^I_\m e^J_\n
- e^I_\n e^J_\m \right)$
 and  $ K =  {\sqrt { \b (1+\g^2_{})}}$ with   $\b$  as an arbitrary function of
space-time coordinates.  We  may   gauge f\/ix the invariance
under boost transformations  by a~convenient choice of $\b$ as follows:

   {\bf Case (i):} A choice of basis is provided by   $\b \equiv \a^2_{}/ (2e)=  -vw/(2e)
\,\he\,  0$
   ($K\, \he\, 0$).
 For this choice,   the $SU(2)$ gauge f\/ields from (\ref{A'}) are:
\begin{gather}
 A^{(1)}_a \, \he \,  {\frac {\g} 2}  \partial^{}_a \ln v + \cos \th   \partial_a^{} \p , \qquad
 A^{(2)}_a \, \he\, 0, \qquad A^{(3)}_a \, \he\,  0   \label{A1}
\end{gather}
  and  equations (\ref{F'}) lead to
  \begin{gather}
  F^{(1)}_{ab}   \,\he\,   2 \partial^{}_{[a} A^{(1)}_{b]}  \,\he\,
- {\frac 2 {r^2_0}}     \S^{23}_{ab}  =  - {\frac {2\g} {r^2_0}}\S^{(1)}_{ab},
\qquad
 F^{(2)}_{ab} \, \he\,  0,   \qquad
  F^{(3)}_{ab} \, \he\,  0.
  \label{F1}
\end{gather}
These relations are invariant under $U(1)$
transformations: $ A^{(1)}_a  \rightarrow A^{(1)}_a - \partial^{}_a \xi$
with~$A^{(2)}_a $ and~$A^{(3)}_a $ unaltered. As we shall show in the
next subsection, these relations
  can be interpreted as the equations of motion of a $SU(2)$
Chern--Simons theory gauge f\/ixed to a $U(1)$ theory.

The $U(1)$ Chern--Simons action for which the f\/irst relation in (\ref{F1}) is
the Euler--Lagrangian equation of motion,  may be written as:
\begin{gather}
 S^{}_1   =  \frac{k}{4\pi} \int^{}_{\D} \e^{abc}_{} A^{}_a \partial^{}_b A^{}_c
+\int^{}_{\D} J^a_{} A^{}_a , \label{CSU1}
\end{gather}
where the non-zero components of
the completely antisymmetric $\e^{abc}_{}$ are given by $ \e^{v \theta \p}_{}= 1$
 and $A^{}_a \equiv A^{(1)}_a$ is the $U(1)$ gauge f\/ield. The external source is given by
 the vector density with upper index $a$ as:
$J^a =   \e^{abc}_{} \S^{(1)}_{bc}/2$. The coupling is directly  proportional to the
horizon area and inversely to the Barbero--Immirzi parameter:
$ k =  \pi r^2_0 /{\g} \equiv  A_{\rm H}/(4\g)$.

    {\bf Case (ii):}  On the other hand, we could make  another  gauge choice where $\b$ is constant
  ($K= \sqrt{\b(1+\g^2_{})} =$  constant) but arbitrary, with gauge f\/ields   given by:
 \begin{gather}
   A^{(1)}_a \,\he\,  \cos \th \partial_a \p,
\qquad A^{(2)}_a \, \he\, -    K
  \left(\cos \d \partial_a \th  -  \sin \d \sin \th  \partial_a \p \right), \nonumber \\
 A^{(3)}_a \, \he\,  -   K  \left( \sin \d \partial_a \th + \cos \d \sin \th  \partial_a \p \right) , \label{A2}
\end{gather}
  where $\cot \d =  \g$.
    The f\/ield strength components  (\ref{F'})  satisfy:
\begin{gather}
  F^{(1)}_{ab}  \equiv  2 \partial^{}_{[ a }A^{(1)}_{b]}
  + 2 A^{(2)}_{[a} A^{(3)}_{b]}   \,\he\,
   - {\frac {2\g} {r^2_0}}  \left[ 1 - \b \left( 1 + \g^2\right) \right]  \S^{(1)}_{ab},\nonumber \\
   F^{(2)}_{ab}  \equiv   2 \partial^{}_{[a} A^{(2)}_{b]}
  +2  A^{(3)}_{[a} A^{(1)}_{b]}  \,\he\, 0, \qquad
  F^{(3)}_{ab}  \equiv  2  \partial^{}_{[a } A^{(3)}_{b]}
  + 2 A^{(1)}_{[a} A^{(2)}_{b]} \,\he\, 0.
  \label{F2}
\end{gather}
  These equations have invariance under   $U(1)$ transformations:
$A^{(1)}_a \rightarrow  A^{(1)}_a - \partial_a^{} \xi$,
$A^{(2)}_a \rightarrow \cos \xi A^{(2)}_a + \sin \xi A^{(3)}_a$ and
$A^{(3)}_a \rightarrow  - \sin \xi A^{(2)}_a  + \cos \xi A^{(3)}_a$.
This ref\/lects that the f\/ield $A^{(1)}_a$ is  a~$U(1)$ gauge f\/ield
and f\/ields $A^{(2)}_a$ and $A^{(3)}_a$ are an $O(2)$ doublet
with $U(1)$ transformations acting as a rotation on them.

Identifying $U(1)$ gauge f\/ield as  $A_a^{} \equiv A^{(1)}_a$ and def\/ining
the  complex vector f\/ields $\phi_a^{}
=  \big( A^{(2)}_a +iA^{(3)}_a\big)/{\sqrt 2}$
and ${\bar \phi}_a^{}
=  \big( A^{(2)}_a -iA^{(3)}_a\big)/{\sqrt 2}$,
    the relations (\ref{F2})
can be recast as:
\begin{gather}
F^{}_{ab} - 2 i {\bar \phi}^{}_{[a} \phi^{}_{b]} \, \he \,  - {\frac {2\g} {r^2_0}}
\left[ 1-\b(1+\g^2_{}) \right] \S^{(1)}_{ab}, \qquad
D^{}_{[a}(A) \phi^{}_{b]} \, \he\, 0,
\qquad D^{}_{[a}(A) {\bar \phi}^{}_{b]} \, \he\, 0,
\label{F2'}
\end{gather}
where the $U(1)$ f\/ield strength is $F^{}_{ab} \equiv 2  \partial^{}_{[a} A^{}_{b]}$
and covariant derivatives of  the charged vector f\/ields   are
$D^{}_a(A)\phi^{}_b \equiv\left( \partial^{}_a +i A^{}_a\right) \phi^{}_b$
and $D^{}_a(A){\bar \phi}^{}_b \equiv \left( \partial^{}_a -i A^{}_a\right){\bar  \phi}^{}_b$
ref\/lecting that $\phi^{}_a$ possesses one unit of $U(1)$ charge.
Now an action principle that would yield (\ref{F2'}) as its equations of motion can be written as:
\begin{gather}
 S^{}_2  =  \frac{k}{4\pi} \int^{}_{\D} \e^{abc}_{} \left[ A^{}_a \partial^{}_b A^{}_c
+ {\bar \phi}^{}_a D^{}_b(A) \phi^{}_c  +   \phi^{}_a D^{}_b(A) {\bar \phi}^{}_c\right]
 +  \int^{}_{\D} J^a_{}A_a^{}, \label{S2}
\end{gather}
where     $ k = \pi r^2_0/  \g   \equiv     A_{\rm H} /(4\g)$ is the coupling  and
  $J^a \equiv  \left[ 1-\b(1+\g^2_{}) \right] \e^{abc}_{} \S^{(1)}_{bc}/2$
is the external source.
There is an arbitrary {\it constant} gauge  parameter $\b$ in the source which can be
  changed by a boost transformation of the original tetrad f\/ields.
  Notice that for  $\b = \left( 1+\g^2\right)^{-1}$, the source vanishes.
The topological f\/ield theory described by action (\ref{S2}) is invariant under
$U(1)$ transformations:  $A^{}_a \rightarrow A^{}_a - \partial^{}_a \xi$, $\phi^{}_a
\rightarrow e^{i\xi}_{} \phi^{}_a$ and ${\bar \phi}^{}_a
\rightarrow e^{-i\xi}_{} {\bar \phi}^{}_a$.

We could interpret the  equations (\ref{F2}) or the equivalent set (\ref{F2'}) alternatively
by taking the combination
$ k =  {\frac {A_{\rm H}} {4 \g\left[ 1 - \b\left( 1+\g^2\right) \right]}} $ to be
the coupling   and $J_{}^a =\e^{abc}_{} \S^{(1)}_{bc}/2 $ as the   source.
This results in a gauge dependent arbitrariness in the  coupling constant, ref\/lected
through the constant parameter $\b$. Boost transformations of the original
gravity f\/ields can be used to change  the value of~$\b$.  In particular, for  $\b = 1/2$, the coupling is $k= \frac{A_{\rm H}}{2\g(1-\g^2_{})}$
and we realize  the gauge theory  discussed in~\cite{per}.

The presence of the arbitrary parameter $\b$ is a  ref\/lection of the ambiguity associated with
 gauge f\/ixing of invariance under boost transformations of the original tetrads $e^I_a$ and connection
f\/ields $\w^{IJ}_a$. Like in any gauge theory, a special choice of gauge f\/ixing only provides
a {\it convenient} description of the theory. No physical quantities should depend on the ambiguity
of gauge f\/ixing. In particular, the Chern--Simons coupling constant is  a physical object.
As we shall see later, physical quantities such as the  quantum horizon entropy depend on
this coupling. This suggests that the coupling $k$ of horizon Chern--Simons theory can not
depend on~$\b$ or any particular value for it. A formulation of the theory that exhibits such
a dependence is suspect. This is to be contrasted with the
dependence on the Barbero--Immirzi parameter~$\g$ which is perfectly possible,
because~$\g$ is not a gauge parameter but a genuine coupling constant (in fact with a
topological origin) of  quantum   gravity.
This   perspective, therefore, picks up the f\/irst interpretation
above for the equations (\ref{F2}) or (\ref{F2'}) as represented by the action (\ref{S2})
with coupling  $k = \pi r^2_0/  \g \equiv A_{\rm H} /(4\g)$  as the correct one.

Notice that the factor $(1+\g^2_{})$ in equations~(\ref{F2}) and~(\ref{F2'})  arises because of
the presence of $\g$ in the gauge f\/ield combinations def\/ined in equations~(\ref{A'}). This
factor does not have any special signif\/icance
as it can be absorbed in the def\/inition of the arbitrary constant boost gauge parameter~$\b$ obtaining a new
boost parameter $\b' = \left( 1+\g^2_{} \right) \b$. Also in equations~(\ref{A'}), we could as well replace~$\g$ by another arbitrary constant~$\l$,  f\/inally leading to the equation~(\ref{F2}) with the factor $(1+\g^2_{})$ replaced by $(1+\l^2_{})$
which again can be absorbed in the arbitrary boost gauge parameter~$\b$ without changing any of the
subsequent discussion. This is to be contrasted with the overall factor of $\g$ in the right-hand
sides of equations~(\ref{F2}) and~(\ref{F2'}), which is  not to be absorbed away in to the boost parameter,
and instead becomes part of the coupling~$k$ of the Chern--Simons theory in~(\ref{S2}).

The boundary topological theory describing the horizon quantum degrees of freedom
and the bulk quantum theory  can be thought of as decoupled from each other except for the sources
of the boundary theory which depend on the bulk quantum f\/ields $\S^{(1)}_{ab}$.
In fact in the bulk theory, $\e^{abc}_{}\S^{(i)}_{bc} /2$ are the canonical conjugate momentum f\/ields
for the Sen--Ashtekar--Barbero--Immirzi $SU(2)$ gauge f\/ields $A^{(i)}_a \equiv \g\w^{0i}_a - \frac{1}{2} \e^{ijk}_{} \w^{jk}_a$. On the other hand, in the boundary
theory described in terms of $U(1)$ Chern--Simons theory,
the f\/ields $(A_\th , A_\p)$ form a  canonical pair. This
allows for the fact that in the classical theory, the boost gauge
f\/ixing of the original gravity f\/ields $(e^I_a,\w^{IJ}_a)$ to obtain the Chern--Simons
boundary theory and that in the bulk theory can be done {\it independently}. In particular,
we could choose $\b \equiv \a^2_{}/(2e) =- vw/(2e)  \,\he\,0$ (or the other choice $\b = {\rm const}$)  for the boundary theory, and make  another  independent {\it convenient}
choice  for the bulk theory, in particular, say the standard time gauge,  so that the
resultant canonical theory in terms of Sen--Ashtekar--Barbero--Immirzi gauge f\/ields in the bulk
can,  at quantum level, lead to  the standard Loop Quantum Gravity theory.

 After these general remarks,   let us now  turn to  discuss how the $U(1)$ invariant  equations~(\ref{F1})
or~(\ref{F2'})  can be
arrived at from a general  $SU(2)$ Chern--Simons theory through a gauge f\/ixing procedure.
This we do in the next subsection.

 We notice that the source  for resultant $U(1)$ gauge theory in either of  the
   cases (i) and  (ii) above is  given in terms of $\S^{(1)}_{ab}
 \equiv \g^{-1}_{} \S^{23}_{ab}$ which is one of the components
 of the $SU(2)$ triplet of   solder forms. An important property to note
 here is that for both these  cases, other two components of this triplet are
 zero on the horizon:
 \begin{gather}
 \S^{(2)}_{\th \p}
 \equiv  {\g}^{-1}_{}\S^{31}_{\th \p} \, \he\,  0\qquad  {\rm and} \qquad \S^{(3)}_{\th
\p} \equiv  {\g}^{-1}_{}\S^{12}_{\th \p} \, \he\,  0,  \label{sourcecond}
\end{gather}
  because $e^1_\th \,\he\, 0$ and $e^1_\p \, \he\, 0$.

   \subsection[Horizon $SU(2)$ Chern-Simons theory]{Horizon $\boldsymbol{SU(2)}$ Chern--Simons theory} \label{II.2}

  The $U(1)$ gauge theories  described by the two sets of  equations
(\ref{F1}) and (\ref{F2}) of the  respective   cases (i) and (ii)
  along with the conditions (\ref{sourcecond}) on the solder forms,  are
 related to   a $SU(2)$
  Chern--Simons theory  through a partial gauge f\/ixing \cite{rkpm}.
  To exhibit this explicitly,   consider the   Chern--Simons action with coupling~$k$:
 \begin{gather}
 S_{\rm CS}  =  {\frac k { 4\pi}} \int_{\D }^{}
\e^{abc}_{} \left( A'^{(i)}_a \partial^{}_b A'^{(i)}_c  +   {\frac  1 3}
 \e^{ijk}_{}  A'^{(i)}_a A'^{(j)}_b A'^{(k)}_c  \right)    +  \int_{\D  }
J'^{(i)a}_{} A'^{(i)}_a , \label{CSAction+}
\end{gather}
where $A'^{(i)}_a$ are the $SU(2)$ gauge f\/ields.
This is a topological f\/ield theory:  the action is independent of the metric of
 three-manifold $\D$.  We take
 the covariantly  conserved  $SU(2)$ triplet of sources, which are   vector densities  with upper index
$a$,   to have  a special form as: $ J'^{(i)a}_{} \equiv\left( J'^{(i)v}_{},
J'^{(i)\theta}_{}, J'^{(i)\p}_{} \right) = \left( J'^{(i)}_{}, 0, 0\right)$.

The   action (\ref{CSAction+})  leads to the Euler--Lagrange equations of motion:
\begin{gather}
  F^{(i)}_{v \theta}(A') \, \he\, 0, \qquad F^{(i)}_{v \p}(A') \, \he\, 0  ,  \qquad
 {\frac k {2\pi}  } F^{(i)}_{\theta \p}(A')   \, \he\,  -    J'^{(i)}_{} ,  \label{CSKruskal1}
\end{gather}
where $F^{(i)}_{ab}(A')$ is the f\/ield strength for the gauge f\/ields $A'^{(i)}_a$.
For  the f\/irst two equations, {\it the most general solution  is given in terms of  the conf\/igurations with
$A'^{(i)}_v$ as  pure gauge}:
$
A'^{(i)}_v = -{\frac 1 2}  \e^{ijk}_{} \left( {\cal O} \partial_v^{} {\cal O}^T_{} \right)^{jk}_{}$,
where ${\cal O}$ is an arbitrary $3 \times 3$ orthogonal matrix,
${\cal O} {\cal O}^T_{} = {\cal O}^T_{} {\cal O} =1$ with $\det {\cal O} =1$.
The other gauge f\/ield components are given in terms of $v$-independent $SU(2)$ gauge potentials
$ B'^{(i)}_\th $  and $B'^{(i)}_\p$ as
$ A'^{(i)}_ {\hat a} = {\cal O}^{ij}_{} B'^{(j)}_{\hat a} -
{\frac 1 2}  \e^{ijk}  \left( {\cal O} \partial_{\hat a}^{} {\cal O}^T_{} \right)^{jk}_{}$ for
${\hat  a} = (\th, \p) $.    Then, since the f\/irst two equations of (\ref{CSKruskal1}) are  identically satisf\/ied,
we are left with the last equation to study:
\begin{gather}
 {\frac k {2\pi}}  F^{(i)}_{\th \p} (A')  =  {\frac k {2\pi}}   {\cal O}^{ij}_{} F^{(j)}_{\th \p}(B')
\, \he \,  -    J'^{(i)}, \label{CSB'}
\end{gather}
 where $F^{(i)}_{\th \p}(B')$ is the $SU(2)$ f\/ield strength constructed from
 gauge f\/ields  $(B'^{(i)}_\th,  B'^{(i)}_\p )$.
  For this set of gauge conf\/igurations, part of the   $SU(2)$ gauge invariance has been f\/ixed
 and, on the spatial slice $\mathbb{S}^2$ of $\D$,  we are now left with
 invariance only  under $v$-independent $SU(2)$ gauge transformations
 of the f\/ields  $B'^{(i)}_\th(\th, \p)$ and $  B'^{(i)}_\p(\th, \p)$.
 Next step  in this construction is to use this gauge
 freedom
to rotate  the triplet   $F^{(i)}_{\th \p}(B')$
to a new f\/ield strength  $F^{(i)}_{\th\p} (B)$  parallel to an internal space  unit vector $u^i_{}(\th ,\p)$.
This can always be achieved through a $v$-independent  gauge
 transformation    $ {\bar {\cal O}}^{ij}_{}(\th, \p)$ with components
 ${\bar {\cal O}}^{i1}(\th, \p) \equiv u^i_{}(\th,\p)$:
\begin{gather}
 F^{(i)}_{\th\p}(B')  =  {\bar  {\cal O}}^{ij}_{} F^{(j)}_{\th \p} (B)
 \equiv  u^i(\th, \p)  F^{(1)}_{\th \p}(B), \nonumber \\
  F^{(1)}_{\th \p} (B) \equiv  u^i_{} F^{(i)}_{\th \p} (B') \ne  0, \qquad F^{(2)}_{\th \p} (B)= 0,  \qquad
F^{(3)}_{\th \p} (B)=0, \label{FB}
\end{gather}
where  the primed and unprimed $B$ gauge f\/ields  are related by a gauge transformation as:
$
B'^{(i)}_{ \hat a} = {\bar {\cal O}}^{ij}_{}
B^{(j)}_{\hat a } - {\frac 1 2}  \e^{ijk} \left( {\bar  {\cal O}} \partial_{\hat a}^{}
{\bar {\cal O}^T_{}}\right)^{jk}_{}
$. We now need to look for the gauge f\/ields $B^{(i)}_{\hat a}$ that solve the  equations~(\ref{FB}).
There are two types of solutions to these equations.
These have been worked out explicitly in the Appendix of~\cite{rkpm}. We shall   summarize the  results
in the following.

  We may parametrize  the internal space unit vector $u^i(\th,\p)$
in terms of two angles $\Theta (\th, \p) $ and $\Phi(\th,
\p)$ as
$u^i_{} (\th, \p) =   {\bar {\cal O}}^{i1}_{}= \left( \cos \Theta , \sin \Theta \cos \Phi , \sin \Theta
\sin \Phi\right)$. Other components of the orthogonal matrix
${\bar {\cal O}}$ in (\ref{FB}) may be written as: ${\bar  {\cal O}}^{i2}_{}=
\cos \chi s^i_{} +\sin \chi  t^i_{} $ and $ { \bar {\cal O}}^{i3}_{} = - \sin \chi s^i_{}
+\cos \chi t^i_{}$ where $\chi (\th, \p)$ is an arbitrary angle and
 $s^i_{}(\th,\p) = \left( - \sin \Theta,  \cos \Theta \cos  \Phi ,  \cos
\Theta \sin \Phi \right) $,  $ t^i_{} (\th,\p) =  \left( 0,  -\sin  \Phi ,  \cos   \Phi \right)$. The angle f\/ields $\Theta(\th, \p)$, $\Phi(\th,\p)$ and
$\chi(\th,\p)$ represent the three independent parameters of the uni-modular orthogonal
transformation matrix ${\bar {\cal O}}$.

Next  we   express   the gauge f\/ields $B'^{(i)}_{\hat a}$, without any loss of generality,
  in terms of their components along and orthogonal to
the unit vector $u^i_{}$ as:
\[
 B'^{(i)}_{ \hat a }  =    u^i B_{\hat a}
 +  f  \partial_{\hat a} u^i   +  g  \e^{ijk} u^j \partial_{\hat a} u^k,
\qquad  {\hat a} = ( \th,\p)
\]
with the f\/ield strength constructed from these as:
\begin{gather}
 F^{(i)}_{{\hat a}{\hat b}}(B')  =
u^i \big( 2 \partial_{[{\hat a}} B_{{\hat b}]} + \left( f^2 + g^2 +2g \right)
\epsilon^{jkl} u^j \partial_{\hat a} u^k \partial_{\hat b} u^l \big)
\nonumber \\
\phantom{F^{(i)}_{{\hat a}{\hat b}}(B')  =}{}
+  2 \partial_{[{\hat a}} u^i\big( (1+g) B_{{\hat b}]}
- \partial_{{\hat b}]} f \big) - 2\epsilon^{ijk} u^j\partial_{[{\hat a}}
u^k \big( f B_{{\hat b}]} + \partial_{{\hat b}]} g\big). \label{FB1}
\end{gather}
Six independent f\/ield degrees  of
freedom in     $B'^{(i)}_{ \hat a}$ are now distributed in $u^i$ (two
independent f\/ields), $B_{\hat a}$ (two f\/ield degrees of freedom) and   two
f\/ields ($f,g$).

 Requiring the f\/ield strength $F^{(i)}_{{\hat a}{\hat b}}(B')$
 in (\ref{FB1}) to satisfy the equations~(\ref{FB}), gives us equations for
various component f\/ields $f$, $g$ and $B^{}_{\hat a}$ which we need to
solve.  There are  two possible solutions to the equations so obtained.
These can be expressed through
two types of gauge f\/ields~$B^{(i)}_{\hat a}$. These gauge f\/ields are related to
$B'^{(i)}_{ \hat a}$ through gauge transformation ${\bar {\cal O}}$ as indicated in~(\ref{FB}).
    We just list these two solutions  here: (a) The f\/irst solution is given by:
    $f=0$, $g=-1$ with  $B^{}_{\hat a}$ as  arbitrary, leading to $
B^{(i)}_{\hat a} =\left( B^{}_{\hat a} +\cos \Theta \partial_{\hat a}^{}\Phi,
0, 0\right)$. Now the conf\/iguration~(\ref{A1}) with its f\/ield strength as in
(\ref{F1}) above can be identif\/ied with this solution for
$B^{}_{\hat a}=0$ and $\Theta = \th$,  $\Phi = \p$  and coupling $k =A_{\rm H}/(4\g)$.
 (b) The second solution
is given by $f=c \cos \delta$, $1+g= c\sin \delta$
and $B^{}_{\hat a} = -\partial^{}_{\hat a} \delta$ with $c$ as a constant
and $\delta(\th, \p)$   arbitrary.
This leads to the gauge conf\/iguration: $ B^{(1)}_{\hat a}
=- \partial_{\hat a}^{} \d+ \cos \Theta \partial_{\hat
a}^{} \Phi$,  $B^{(2)}_{\hat a} = c \left( \cos \d \partial_{\hat a}^{}
\Theta - \sin \d  \sin \Theta \partial^{}_{\hat a}\Phi\right)$  and $
B^{(3)}_{\hat a} =   c \left( \sin \d \partial_{\hat a}^{} \Theta + \cos \d
\sin \Theta \partial^{}_{\hat a}\Phi\right)$.   Now the conf\/iguration
  (\ref{A2}) with its f\/ield strength
components satisfying the equations (\ref{F2})  can be identif\/ied with
this solution for    $c=-K$ and $\Theta = \th$,  $\Phi=\p $ and $\d$ as a constant.
 Further, for $c=0$ and  constant $\delta $, this solution coincides with the f\/irst solution
(a) above for $B^{}_{\hat a} =0$.

Finally, we may rewrite the  starting $SU(2)$ gauge conf\/igurations $A'^i_a$ of~(\ref{CSKruskal1}),
 for both these cases,   as:
$ A'^{(i)}_v = -{\frac 1 2} \e^{ijk}_{} \big( {\cal O'} \partial_v^{}
{\cal O'}^T_{} \big)^{jk}_{}$, $A'^{(i)}_ {\hat a} = {\cal O'}^{ij}_{}
B^{(j)}_{\hat a} -  {\frac 1 2} \e^{ijk}_{} \big( {\cal O'} \partial_{\hat
a}^{} {\cal O'}^T_{} \big)^{jk}_{}$ where ${\cal O'}$ is the product of gauge
transformation matrices introduced in (\ref{CSB'}) and (\ref{FB}): ${\cal O'}= {\cal O} {\bar {\cal O}}$.
The f\/irst two equations of (\ref{CSKruskal1}) are identically satisf\/ied
and   the last equation
 becomes
 \begin{gather*}
 {\frac k {2 \pi}  }  F^{(i)}_{\th \p} (A')  =
{\frac k {2 \pi} }  {\cal O'}^{ij}_{} F^{(j)}_{\th \p}(B)   \, \he \,  -
J'^{(i)}  \equiv  -   {\cal O'}^{ij} J^{(j)} ,  
\end{gather*}
where now
from (\ref{FB}),  $F^{(i)}_{\th \p}(B) = \big( F^{(1)}_{\th \p} (B), 0,0
\big)$, which implies for  the sources
$J^{(i)} = \left( J, 0,0 \right)$. Thus, this
gauge f\/ixing procedure leads to
   a theory described in terms of
f\/ields   $B^{(i)}_{\hat a}$ with a left over
invariance only under $U(1)$ gauge transformations.

 This completes our discussion of
how the horizon properties  can be described by a $SU(2)$ Chern--Simons
gauge theory or equivalently, by a gauge f\/ixed version with only  $U(1)$
invariance.

\subsection{Isolated horizons} \label{II.3}

In order to def\/ine horizons in a manner decoupled from the bulk,   a generalized notion of  Isolated
Horizon (IH), as a quasi-local replacement of the event horizon of a black hole,  has been
developed by Ashtekar et al.~\cite{ash1}. This is done  by ascribing attributes
   which are  def\/ined on the horizon intrinsically  through a set of quasi-local boundary conditions
without  reference to  any assumptions
like stationarity such that the horizon is isolated
in a precise sense. This permits us to describe a black hole in equilibrium with
a dynamical exterior region. An IH  is def\/ined to
be a null surface, with topology $\mathbb{S}^2 \times \mathbb{R}$, which is non-expanding and shear-free.
The va\-rious geometric quantities on such a horizon are seen to satisfy $U(1)$  Chern--Simons
equations of motion~\cite{ash2}:
\begin{gather}
  F^{}_{r \th}  = 0, \qquad F^{}_{r \p}  = 0 , \qquad F^{}_{\th \p}  = - \frac{2\pi}{k}
\S^{(1)}_{\th \p}, \label{IHCS}
\end{gather}
where $r$, $\th$ and $\p$ are the coordinates on the horizon
and $k =  A_{\rm H}^{} /(4\g)$ with $A_{\rm H}^{}$ as the  horizon~area, $\g$ is the Barbero--Immirzi parameter,
$F_{ab}$ is the f\/ield strength of $U(1)$ gauge f\/ield.
The source $\S^{(1)}_{\th \p}$ is one component of the $SU(2)$ triplet of solder  forms $\S^{(i)}_{\th \p}
\equiv \g^{-1}_{} \e^{ijk}_{}e^j_\th e^k_\p$, in the direction of the subgroup $U(1)$.
There is  another fact which is not some times  stated expli\-cit\-ly. The horizon boundary conditions,
which lead to the equations (\ref{IHCS}), also further imply the following constraints for the components
of the triplet of  solder  forms in the internal space directions orthogonal to the $U(1)$ direction:
\begin{gather}
  \S^{(2)}_{\th \p} = 0 , \qquad \S^{(3)}_{\th \p} = 0. \label{addcond}
\end{gather}

Now this is exactly the same situation as we came across for the Schwarzschild hole
in Section~\ref{II.1} above.
Just like there, the equations~(\ref{IHCS}) and~(\ref{addcond}) really
describe a $SU(2)$ Chern--Simons theory partially  gauge f\/ixed to leave  only a leftover~$U(1)$ invariance.

Thus, as in the case of  the Schwarzschild hole, the degrees of freedom of the more general
Isolated Horizon  are
also described by a quantum $SU(2)$ Chern--Simons gauge theory with specif\/ic sources given in terms of the
solder f\/ields.  An equivalent description is provided by a~gauge f\/ixed version of this theory
in terms of the quantum $U(1)$ Chern--Simons theory represented by the   operator
constraint~(\ref{IHCS}),
{\it but with the physical states satisfying additional conditions  which are the quantum analogues of
the classical constraints~\eqref{addcond}}.
Horizon properties like the   entropy can be calculated in either version with the same  consequences.
 We shall review these calculations in the following.

\section{Micro-canonical entropy} \label{III}

Over last several decades, many   authors have developed methods based on   $SU(2)$
gauge theory to count the micro-states associated
with a two-dimensional surface. Smolin was  f\/irst to explore the use of $SU(2)$ Chern--Simons theory
induced on a  boundary satisfying self-dual boundary conditions in Euclidean gravity~\cite{smo}. He
also demonstrated that such a boundary theory obeys the Bekenstein bound.
  Krasnov   applied
these ideas to the black hole horizon and used the ensemble of quantum states of  $SU(2)$
Chern--Simons theory associated with  the spin assignments of the punctures on the surface
to count the boundary degrees of freedom
and reproduced  an area law for the entropy \cite{kras}. In this f\/irst application of
$SU(2)$ Chern--Simons theory to black hole entropy,  the gauge coupling was
taken to be proportional to the horizon area and also inversely proportional to the Barbero--Immirzi
parameter.
Assuming that the quantum states of a f\/luctuating
black hole horizon to be governed by the properties of intersections of
knots carrying $SU(2)$ spins with the two-dimensional surface,
Rovelli also developed a counting procedure which
again yielded  an area law for the entropy \cite{rov}. In the general context of Isolated Horizons,
it was  the pioneering work of Ashtekar, Baez, Corichi and Krasnov~\cite{ash2}  which
studied $SU(2)$ Chern--Simons theory as the boundary theory and  the area law for entropy
was again reproduced. This  was further developed in~\cite{kmplb, kmprl,dkmprd, krprd}
which extensively exploited the deep
connection between  the three dimensional Chern--Simons theory   and the   conformal f\/ield theories
in two dimensions.
This framework provided a  method  to calculate corrections beyond  the area law for
micro-canonical entropy  of large black holes.
In particular, it is more than ten years now when    a leading correction given by the logarithm of
horizon area with a def\/inite coef\/f\/icient of~$-3/2$  followed by sub-leading terms
containing  a constant and inverse powers of area were f\/irst obtained~\cite{kmprl}:
\[
 S^{}_{\rm bh}  =  S^{}_{\rm BH}  -  \frac{3}{2}  \ln S^{}_{\rm BH}  +  {\rm const}
 +  {\cal O}\big(S^{-1}_{\rm BH}\big),
\]
where $S^{}_{\rm BH}= A_{\rm H}/(4 \ell^2_P)$ is the Bekenstein--Hawking entropy
given in terms of horizon area~$A_{\rm H}$.  The  corrections
due to the non-perturbative ef\/fects represented by the discrete quantum geo\-metry are f\/inite.
These may be contrasted with those obtained in Euclidean path integral formulation from the
graviton and other quantum matter f\/luctuations around the hole back ground which
depend on the renormalization scale~\cite{ms}.

\subsection[Horizon entropy from the  $SU(2)$ Chern-Simons theory]{Horizon entropy from the  $\boldsymbol{SU(2)}$ Chern--Simons theory} \label{III.1}

In this subsection,  we shall survey the general framework developed in~\cite{kmplb,kmprl}
 for studying  the horizon properties  in the $SU(2)$ Chern--Simons formulation.
   An important ingredient in counting the  horizon  micro-states  is the
 fact   \cite{witten, wbt} that Chern--Simons theory on a three-manifold with boundary can be
completely described by
 the properties of a gauged  Wess--Zumino conformal theory on that two dimensional boundary.
     Starting with  the pioneering work of Witten
 leading to Jones polynomials~\cite{witten}, this relationship has been extensively used to study
  Chern--Simons
 theories. This includes methods to solve the Chern--Simons theories explicitly and exactly
 and also to construct
three-manifold  invariants from generalized knot/link invariants in these theories~\cite{kaultop}.

In  the  LQG, the Hilbert space of canonical quantum gravity is described by spin networks with
Wilson line operators  carrying $SU(2)$ representations  (spin $j=  1/2, 1, 3/2,\dots$) living
on the edges of the graph.  Sources of the boundary $SU(2)$ Chern--Simons theory (with coupling $k=A_{\rm H}/(4\g)$)
 describing the horizon properties are
given in terms of the solder forms which are quantum f\/ields of the bulk theory.
These have distributional  support at the   {\it punctures}  at which the bulk
spin network edges impinge on the horizon. Given the relationship of Chern--Simons theo\-ry
 and the two dimensional conformal f\/ield theory mentioned above, {\it the Hilbert space of states of $SU(2)$
Chern--Simons theory with coupling $k$ on
a three-manifold $\mathbb{S}^2_{} \times \mathbb{R}$ $($horizon$)$ is completely characterized by the conformal blocks
of the $SU(2)_k$ Wess--Zumino conformal
theo\-ry on an  $\mathbb{S}^2$ with   punctures ${\cal P} \equiv\left\{1, 2,\dots,p\right\}$  where each
puncture carries spin
$j^{}_i = 1/2,1,3/2, \dots,k/2$.}

\begin{figure}[t]
\centerline{\includegraphics{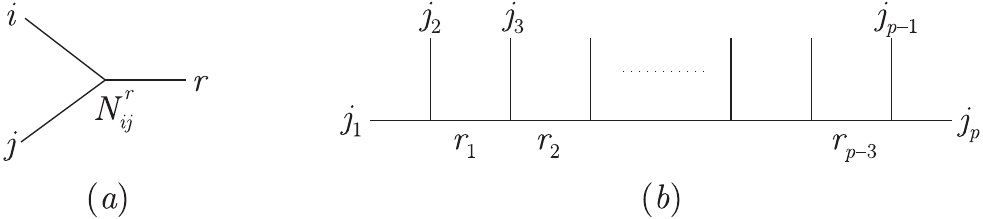}}

\caption{Diagrammatic representation for $(a)$ the fussion matrix $N^{r}_{ij}$  and $(b)$
for the composition rule~(\ref{NP1}) for spins $j^{}_1, j^{}_2, j^{}_3, \dots,j^{}_p$.}\label{Kaul-Fig1}
\end{figure}

$SU(2)_k$ conformal f\/ield theory is described in terms of primary f\/ields $\p^{}_j$ with  spin values
cut of\/f by the maximum value $k/2$, $j=0,1/2,1, \dots,k/2$. The composition rule for two spin~$j$ and~$j'$
representations
is modif\/ied from that in the  corresponding ordinary $SU(2)$  as: $(j)\otimes(j')
= (|j-j'|) \oplus (|j-j'| +1) \oplus (|j-j'|+2) \oplus \cdots \oplus ( \min ( j+j',  k -j -j'))$.
We may rewrite this composition
law for the primary f\/ields $[ \phi^{}_i]$
and $[\phi^{}_{j}]$   as: $[\phi^{}_i] \otimes [\phi^{}_j]$
$= \sum^{}_r N^{r}_{ij} [\phi^{}_r]$  in terms of the fusion matrices~$N^{r}_{ij}$ whose
elements have values $1$ or $0$, depending on whether the primary f\/ield $[\phi^{}_r]$ is allowed or
not in the product.  Representing the fusion matrix~$N^{r}_{ij}$
diagrammatically as in Fig.~\ref{Kaul-Fig1}$(a)$, the composition of $p$ primary f\/ields in  spin representations
$j^{}_1, j^{}_2, j^{}_3, \dots, j^{}_p$ can be depicted by the diagram in Fig.~\ref{Kaul-Fig1}$(b)$.
Then the total number of conformal blocks    with spin $j^{}_1, j^{}_2,\dots,j^{}_p$
on the  external lines
(associated  with the $p$ punctures on $\mathbb{S}^2$) and spins $ r^{}_1,r^{}_2,\dots, r^{}_{p-3}$
on the internal lines
in this composition diagram is the product of $(p-2)$ factors of  fusion matrix as  given by:
\begin{gather}
 {\cal N}^{}_{\cal P}  =  \sum^{}_{\{r^{}_i\}} N^{r^{}_1}_{j^{}_1 j^{}_2}
N^{r^{}_2}_{r^{}_1 j^{}_3}
 N^{r^{}_3}_{r^{}_2 j^{}_4} \cdots N^{j^{}_p}_{r^{}_{p-3} j^{}_{p-1}}. \label{NP1}
\end{gather}
There is a remarkable result due to Verlinde which states that the fusion matrices $(N_i)^{r}_j \equiv
N^{r}_{ij}$ of a~conformal f\/ield theory are diagonalised by the unitary duality
matrices $S$ associated with modular
transformation $\t \rightarrow -1/\t$ of the torus. This fact immediately leads to the Verlinde
formula which expresses the components of the fusion matrix in terms of those of this~$S$ matrix~\cite{ver}:
\begin{gather}
 N^{r}_{ij} = \sum^{}_s \frac{S^{}_{is} S^{}_{js} S^{\dagger r}_s  } {S^{}_{0s}}. \label{N}
\end{gather}
For the $SU(2)_k^{}$
Wess--Zumino conformal theory,  the duality matrix  $S$  is explicitly given by
\begin{gather}
 S^{}_{ij} = \sqrt{ \frac{2}{k+2} } \sin \left( \frac{(2i+1)(2j+1)\pi}{k+2} \right) ,\label{S}
\end{gather}
where $i= 0,1/2,1, \dots,  k/2$ and  $ j= 0,1/2,1,\dots,  k/2$ are the spin labels.

The fusion rules and Verlinde formula    above  were f\/irst obtained in the conformal theory context.
{\it It is also possible to derive these results directly in the Chern--Simons theory,
using only the gauge theory techniques  without taking  recourse to the
conformal f\/ield theory.   This has been done  in  the paper of Blau and Thompson in~{\rm \cite{wbt}}.}
This paper also discusses how the   Chern--Simons theory based on a compact gauge group $G$
 can be abelianized  to a topological
f\/ield theory based on the maximal torus~$T$  of~$G$.  In particular, for the $SU(2)$ Chern--Simons
theory, this framework describes the gauge f\/ixing to the maximal torus of $SU(2)$ which is
its~$U(1)$ subgroup.

Now, the formula (\ref{NP1}) for the   number of conformal blocks ${\cal N}^{}_{\cal P}$
for the set of punctures ${\cal P}$  can be rewritten, using (\ref{N}) and unitarity of the matrix~$S$, as:
\begin{gather*}
{\cal N}^{}_{\cal P} =  \sum^{k/2}_{r=0} \frac{S^{}_{j^{}_1 r} S^{}_{j^{}_2 r}\cdots
 S^{}_{j^{}_p r}}  {\left( S^{}_{0 r}\right)^{p-2}_{}}, 
\end{gather*}
which further, using the explicit formula for the duality matrix (\ref{S}),  leads to \cite{wbt, ver, kmplb, kmprl}:
\begin{gather}
 {\cal N}^{}_{\cal P}  =  \frac{2}{k+2}  \sum^{k/2}_{r=0}  \frac{\prod\limits^p_{l=1}
\sin\left(\frac{(2j^{}_l+1)(2r+1)\pi}{k+2}\right)}{\left[ \sin\left(\frac{(2r+1)\pi}{k+2}
\right)\right]^{p-2}_{} }.
\label{NP3}
\end{gather}
{\it This master formula just counts the number of ways   $p$
   primary fields in spin
$j^{}_1, j^{}_2,\dots,j^{}_p$
representations associated with the $p$ punctures on $\mathbb{S}^2_{}$ of horizon can be composed into
$SU(2)$ singlets}. Notice the presence of  combination $k+2$ in this formula. This   just
  ref\/lects   the fact that  the
{\it effective}  coupling constant of   quantum $SU(2)$ Chern--Simons theory is $k+2$
instead of its classical value~$k$.

Now, the horizon entropy is given by counting the micro-states by summing ${\cal N}^{}_{\cal P}$
over all possible  sets of punctures  and then taking its logarithm:
\begin{gather}
  {\cal N}^{}_{\rm H}  =  \sum^{}_{\{{\cal P}\}} {\cal N}^{}_{\cal P} , \qquad S^{}_{\rm H}  =
  \ln {\cal N}^{}_{\rm H}       \label{en1}
\end{gather}
for a f\/ixed  horizon area $A_{\rm H}$ (or more accurately with nearby area values in a suf\/f\/iciently narrow
range with this f\/ixed mid point  value). In   LQG,   area for  a punctured $\mathbb{S}^2$,
  with the spins $j^{}_1, j^{}_2, \dots, j^{}_p$ on the $p$ punctures  is given by
\cite{rovbk}:
\begin{gather}
 A_{\rm H}  =  8\pi \g  \sum^{ }_{l=1,2,\dots,p}\sqrt{ j^{}_l(j^{}_l +1) } \label{area1}
\end{gather}
in the units where the Planck length $\ell^{}_P=1$. Here  $\g$ is the Barbero--Immirzi parameter.

A straightforward  reorganization of the master formula (\ref{NP3}), through a redef\/inition of the dummy
variables and using the fact that the product in this formula can be written as a multiple sum,
leads to an alternative equivalent expression as \cite{kmplb}:
\begin{gather}
{\cal N}^{}_{\cal P} =  \frac{2}{k+2}  \sum^{k+1}_{\ell=1,2, \dots} \sin^2_{} \frac{\th^{}_\ell}{2}
\left\{\sum^{j^{}_1}_{m^{}_1=-j^{}_1} \cdots \sum^{j^{}_p}_{m^{}_p =-j^{}_p}
\exp\left[i \th^{}_\ell  \left( m^{}_1+m^{}_2+\cdots +m^{}_p\right)\right]\right\} \label{NP4}
\end{gather}
with $\th^{}_\ell \equiv \frac{2\pi l}{k+2}$. Now, we use the representation for a periodic Kronecker delta,
with period $k+2$:
\begin{gather*}
  {\bar {\delta}}^{}_{m^{}_1+m^{}_2 +\cdots +m^{}_p ,   m}  \equiv
 \frac{1}{k+2}   \sum^{k+1}_{\ell=0} \exp\left[ i\th^{}_\ell \left( m^{}_1+m^{}_2 +\cdots + m^{}_p -m
\right)\right] .
\end{gather*}
Expanding the $\sin^2  \frac{\th^{}_\ell}{2}$ factor in the formula (\ref{NP4}) and after an interchange of the summations,
this formula can   be recast as~\cite{kmplb}:
\begin{gather}
{\cal N}^{}_{\cal P}= \!\sum^{j^{}_1}_{m^{}_1=-j^{}_1} \!\cdots \! \sum^{j^{}_p}_{m^{}_p=-j^{}_p}
\!\!\left({\bar {\delta}}^{}_{m^{}_1+m^{}_2+\cdots +m^{}_p,0} -\frac{1}{2} {\bar { \delta}}^{}_{m^{}_1
+m^{}_2 +\cdots +m^{}_p,1}  - \frac{1}{2}  {\bar {\delta}}^{}_{m^{}_1+m^{}_2 +\cdots +m^{}_p, -1}
\right).\!\!\!\! \label{NP5}
\end{gather}
The various terms here have specif\/ic special interpretations \cite{dkmprd}:
The f\/irst term just counts the total number of ways the `magnetic'
  quantum number $m$ of the spin $j^{}_1, j^{}_2,\dots, j^{}_p$  assignments  on the $p$ punctures can be
added to yield   total $m^{}_{\rm tot} =\sum\limits^{p}_{l= 1} m^{}_l=0$  modulo $k+2$.
This sum over counts the total number of singlets
($j^{}_{\rm tot} =0$)  in the
composition of primary f\/ields with spins  $j^{}_1, j^{}_2, \dots, j^{}_p$,  because it also includes
those states
with $m^{}_{\rm tot} =0$ coming from conf\/igurations  with total spin  $j^{}_{\rm tot} =1, 2, \dots $ in the product
representation $\otimes^{p}_{l=1}  (j^{}_l)\equiv ( j^{}_1)\otimes ( j^{}_2)\otimes \cdots \otimes (j^{}_p)$.
Such states are always accompanied by those with $m^{}_{\rm tot}=\pm 1$
in the product $\otimes^p_{l=1} ( j^{}_l)$.  Hence these  can be
counted by enumerating
  the number of  ways the $ m$ quantum numbers of the spin representations $j^{}_1, j^{}_2, \dots,
j^{}_p$
add up to $m^{}_{\rm tot}= \sum\limits^p_l m^{}_l= +1$ (modulo $k+2$)  or $m^{}_{\rm tot}=-1$  (modulo $k+2)$. Note,
these two numbers are equal which makes the  last two
terms  in~(\ref{NP5})  equal. Hence with the normalization  factor $1/2$  in
 each of them,  these two terms   precisely subtract the number of  extra states  so that
{\it formula \eqref{NP5} counts   exactly  the number of singlet states in
the product $\otimes^p_{l=1} (j^{}_l)$.}

Presence of the periodic Kronecker  deltas ${\bar {\delta}}_{m,n}$
in (\ref{NP5}) distinguishes this formula of the $SU(2)_k$  Wess--Zumino
conformal f\/ield theory from the  corresponding  group theory formula  for  $SU(2)$ with ordinary
Kronecker deltas $\delta_{m,n}$.  In the  large limit $ k$ $(k\gg 1)$, the periodic
Kronecker delta ${\bar {\delta}}_{m,n}$ can be approximated by the ordinary Kronecker delta $\delta^{}_{m,n}$;
hence in this limit, the equation (\ref{NP5}) leads to the ordinary $SU(2)$ group  theoretic
formula for counting   singlets in the composite representation $ \otimes^{p}_{l=1}  (j^{}_l)$.

The master formula (\ref{NP3}) along with its equivalent representations (\ref{NP4}) and (\ref{NP5}) and the
entropy formula
(\ref{en1})  are exact and provide a general framework, f\/irst set up in \cite{kmplb, kmprl},  for study
of horizon entropy. For  large
horizons,   suitable approximate methods have been adopted to  extract interesting results from these
equations. For   f\/ixed large values of~$p$ and  the horizon area~$A_{\rm H}$, it is clear that the largest
contribution to the degeneracy
of horizon states  will come from {\it low} values of the spins $j_i$ assigned to the punctures.
For computational simplicity, let us put spin~$1/2$ representations on all the puncture sites on $\mathbb{S}^2$.
The dimension of the associated Hilbert space in this case can be readily evaluated.   To obtain the
leading behaviour for large~$k$ $\big(=\frac{A_{\rm H}}{4\g}\big)$,
 the state counting can as well be done using ordinary $SU(2)$
rules. It is straight forward to check that, for the case with spin~$1/2$ on all the punctures, this yields:
 \begin{gather}
 {\cal N}_{\cal P} =    \begin{pmatrix}
                         p  \\ p /2
                        \end{pmatrix}
                  -    \begin{pmatrix}
                         p  \\ (p /2-1)
                         \end{pmatrix} .           \label{NP1/2}
\end{gather}
The f\/irst term here follows from the f\/irst term of (\ref{NP5}) in the limit of large  $k$ $(k\gg 1)$
 and  simply  counts the number of ways $m^{}_i =\pm 1/2$ assignments can be put on $p$ (even) punctures
 such that  $\sum m^{}_i =0$. The second term of (\ref{NP1/2}) similarly follows from the second and third
 terms of~(\ref{NP5}), counting the number of ways assignments $m^{}_i =\pm 1/2$ are placed on the
 punctures such that  their sum is $+1$ or $-1$. The dif\/ference of these two terms counts the number of
 $SU(2)$ singlets in the product of $p$ spin $1/2$ representations. The expression (\ref{NP1/2})
 is also equal to   $n$th number in the Catalan series $C^{}_n = \frac{(2n)!}{(n+1)! n!}$
for $p=2n$. For  large $p$, using Stirling formula, this leads to~\cite{kmprl, dkmprd}:
\[
 {\cal N}^{}_{\cal P}  =  C  \frac{2^p}{p^{3/2}}\left\{ 1+ {\cal O}
\left(\frac{1}{p}\right) \right\},
\]
where $C$ is a $p$ independent  numerical constant\footnote{It is also possible to count the number ${\cal N}^{(j)}_{\cal P}$  of ways spin  $1/2$ representations on
$p$ punctures can be composed to yield, instead of the singlets, net spin $j$ representations. This follows
from  straight
forward application of the techniques developed in~\cite{kmprl} which yield, for   $j\ll  k$ and
  large $k$ and $p$:
${\cal N}^{(j)}_{\cal P} \sim 2^{p+2}_{} \left[ F(p) - F(p+2) \right]$ where
$F(p) \approx \frac{1}{\pi} \int^{\pi }_0 dx  \left( \frac{\sin[(2j+1)x]}{\sin x} \right)   \cos^p x$.
For $j\ll p$, this integral can be evaluated to be $F(p) \sim \frac{(2j+1)}{\sqrt p} \left\{ 1 + {\cal O}(\frac{1}{p})
\right\}$ which
 f\/inally leads to the formula: 
 ${\cal N}^{(j)}_{\cal P} \sim
\frac{(2j+1) 2^p  }{p^{3/2}}
\left\{ 1+ {\cal O} \left(\frac{1}{p}\right) \right\}$
[Kaul R.K.,  Kalyana Rama S., unpublished].
 The angular momentum of a rotating black hole is to be def\/ined
with respect to the global rotation properties of the space-time at spatial inf\/inity.  In the LQG,
internal gauge group $SU(2)$ is asymptotically linked with  these global rotations leading
to an identif\/ication of angular momentum with internal spin  at this boundary. However, from the point of view of
the horizon boundary theory, angular momentum, like all other properties of the black hole, has to be
described completely in terms of the horizon attributes  which are codif\/ied in the   topological properties
 of the punctures carrying  the internal spins ${\vec J}_i$  on them. Such an angular momentum operator has to obey the standard
  $SO(3)$ algebra $\big[ J^{(l)},  J^{(m)}\big] = i \e^{lmn} J^{(n)}$.  An operator with these properties
  is the  total spin ${\vec J} = \sum\limits^p_{i=1} {\vec J}_i$. This perspective,
  therefore, suggests that ${\cal N}^{(j)}_{\cal P}$ above represents the degeneracy associated with the horizon states of a rotating   hole
  with quantum angular momentum given by the eigenvalues $ j(j+1)$ of  ${\vec J}. {\vec J}$
   and  its logarithm  may be interpreted as the entropy.}.
Instead, if we place spin $1$ on all the $p$ punctures,   this number is \cite{krprd}:
$
{\cal N}^{}_{\cal P}  =  C  \frac{3^p}{p^{3/2}}\left\{ 1+ {\cal O}\left(\frac{1}{p}
\right) \right\}
$. More generally, for the case where all the punctures carry the same low spin value
$r$ ($r\ll p$ and $r \ll  k$) \cite{krprd}, we have\footnote{Note that for $r$ with half-integer
values, the number of punctures $p$ has to be even in order to get a non-zero number of
net $j^{}_{\rm tot} =0$ states.}:
\[
{\cal N}^{}_{\cal P}  =   C  \frac{(2r+1)^p_{}}{p^{3/2}}
\left\{ 1+ {\cal O}\left(\frac{1}{p}\right) \right\}.
\]
 The corresponding entropy for all these cases is:
\[
S^{}_{\rm H} = \ln {\cal N}^{}_{\cal P}  =  p \ln(2r+1)  -  \frac{3}{2}  \ln p  +  {\cal O}\big( p^0\big).
\]

Now, the area of a two-surface with  Wilson lines, carrying common spin $r$ representation,
impinging on it at $p$ punctures,  is given by (\ref{area1}) as:
$A_{\rm H} = 8\pi\g    p \sqrt{r(r+1)}$. Inverting this to write $p$ in terms of the horizon
area $A_{\rm H}$, $p = A_{\rm H}  \left[8\pi\g     \sqrt{r(r+1)}\right]^{-1}_{}$,
 leads to the entropy formula for large area as \cite{kmprl}:
 \begin{gather}
  S^{}_{\rm H} =   \frac{A_{\rm H}}{4  }  -   \frac{3}{2}  \ln\left( \frac{A_{\rm H}}{4  } \right)
  + {\rm const}  + {\cal O}\big( A^{-1}_{\rm H}\big), \label{en3}
\end{gather}
where the Barbero--Immirzi parameter is f\/ixed to be $\g =\frac{\ln(2r+1)}{2\pi \sqrt{r(r+1)}}$
to match  the linear area term with   the Bekenstein--Hawking law.
The linear area term for $r=1/2$ had been already obtained in~\cite{ash2}. The framework of~\cite{kmplb, kmprl}
provides a systematic procedure of deriving corrections beyond this leading term.
An important point to note
here  is that {\it the sub-leading logarithmic correction is independent of $\g$  and is
insensitive to the   value
of spin $r$ $(r\ne0)$ placed on the punctures}.  But, the coef\/f\/icient of  leading area term   does
depend on  spin value
  $r$  and hence  $\g$ changes as we change $r$.
   The general form of the formula (\ref{en3})
is robust enough,  though the coef\/f\/icient of the linear area term and thereby the value of the
Barbero--Immirzi parameter are not.

 We shall close this discussion  with one last remark. The  derivation of the leading    terms
of the asymptotic entropy formula  (\ref{en3}) does not require the full force of the
conformal f\/ield theory. It was already pointed out in \cite{dkmprd, krprd} and as
has been  seen above that,  instead of  full $SU(2)_k$  composition rules,
use of  {\it ordinary} $SU(2)$
 rules (corresponding to large $k$ $\big(=\frac{A_{\rm H}}{4\g}\big)$ limit)   is good
enough  for the relevant counting   to yield the f\/irst
two terms of      entropy formula (\ref{en3}).
These results have been  re-derived    in \cite{liv1} where ordinary $SU(2)$
composition over the spins   is computed through an
equivalent representation in terms of a random walk modif\/ied
  with a mirror at origin. In this study, the coef\/f\/icient $-3/2$ for the logarithmic
  correction is again
  conf\/irmed   and    this coef\/f\/icient is interpreted as
  a ref\/lection of entanglement
  between parts of the horizon.
However, the terms beyond the logarithmic correction are sensitive to the
details of   conformal theory resulting in ef\/fects which are more pronounced for smaller $k$.
Thus  two ways of
counting would show dif\/ferences
in these terms.

\subsection[Improved value of the Barbero-Immirzi parameter]{Improved value of the Barbero--Immirzi parameter} \label{III.2}

Early  calculations   of the entropy, as discussed above, were done by
taking a {\it common   low value}
of spin, $1/2$ or $1$,  $\dots$,   on {\it all}  the punctures.
Soon it was realized that this approximation needs to be improved \cite{meiss}.
In fact, maximized entropy, subject to holding the horizon area f\/ixed within
a suf\/f\/iciently narrow range $\left[A_{\rm H} -\e, A_{\rm H} +\e\right]$,
is obtained for conf\/igurations with dif\/ferent spin values  $1/2$, $1$, $3/2$, $2$, \dots\
distributed over the punctures in a def\/inite way. The relevant conf\/igurations are
those where fraction $f^{}_j$ of $p$ punctures
carrying spin $j$ representation is given by the probability distribution
 \cite{meiss,gm}:
\begin{gather}
 f^{}_j \equiv   \frac{n^{}_j}{p}  =    (2j+1) \exp \left(-\lambda
\sqrt{j(j+1)}\right).    \label{fj}
\end{gather}
 From this,  we have
\begin{gather}
 \sum_{j=1/2}^{\infty} f^{}_j  \equiv  \sum^{\infty }_{j=1/2} (2j+1) \exp \left(-\lambda
\sqrt{j(j+1)}\right)   =   1, \label{fj1}
\end{gather}
which when   solved numerically  yields
 $\lambda \approx 1.72$.  Using this value of $\lambda$, the   distribution (\ref{fj})
implies that the   conf\/igurations of interest contain   fractions
$f^{}_j  \approx 0.45,   0.26,  0.14,   0.07,    0.04,   0.02, \dots$ of total number of $p$
punctures with spins $j = 1/2, 1,    3/2,  2,  5/2,  3, \dots$ respectively.  Notice
that the low spin values
have higher occupancies; those for higher spin fall of\/f rapidly.

This  improved counting of   net $SU(2)$ spin zero  conf\/igurations
does not change  the ge\-ne\-ral form of the asymptotic
entropy formula~(\ref{en3}).
 In particular, as we shall see below,  the coef\/f\/icient~$-3/2$ of the logarithmic term is unaltered\footnote{The
computations in \cite{meiss, gm} were   originally done in the $U(1)$
  framework without imposing the quantum analogues of the additional constraints (\ref{addcond})
and  logarithmic correction  turned out to be with a coef\/f\/icient~$-1/2$. Same $ - (\ln A_{\rm H})/2$
correction was earlier obtained through  the counting rules analogous to those in $U(1)$ theory
in~\cite{dkmprd}.  However, when done with care including these additional constraints, this
coef\/f\/icient  gets corrected  to~$-3/2$. See the discussion in Section~\ref{III.3}.}.
Only change is in  the coef\/f\/icient of the leading area term.

 The equations (\ref{fj}) and (\ref{fj1})  have been derived in \cite{meiss,gm} by maximizing the entropy subject to
 the f\/ixed area constraint without any further conditions   so that at each puncture carrying spin
 $j$, there are  $2j+1$ possible degrees of freedom. Imposing further constraints, in the $U(1)$ formulation,
 so that the net $U(1)$ charge on all the punctures is zero, modif\/ies these equations  only marginally \cite{gm}.
  For the case of $SU(2)$ formulation where
  the spins on all the punctures have to add up to form $SU(2)$ singlets,
  the corresponding equations also have  modif\/ications which, as we shall see below,
  are suppressed as powers of inverse area.

 In the $SU(2)$ Chern--Simons formulation, for a conf\/iguration where spin $j$ lives on $n_j$ punctures,
 the degeneracy formula (\ref{NP3}) for the set of punctures $\{{\cal P} \}$ with  occupancy numbers $\{n^{}_j\}$
 can be recast as:
 \begin{gather}
 {\cal N} (\{ n^{}_j\}) = \left[ \frac{\left(\sum_j n_j\right)!}{\prod_j  n^{}_j !}\right] \frac{2}{k+2}
 \sum^{k+1}_{\ell = 1,2,\dots }
\sin^2\frac{\th_\ell}{2}  \prod_j \left[ \frac{\sin\left( \frac{(2j+1)\th^{}_\ell}
{2}\right)}{\sin \frac{\th_\ell}{2}} \right]^{n_j}, \label{degen}
\end{gather}
where $\th_\ell \equiv \frac{2 \pi\ell }{k+2}$ and $\sum_j n_j^{} =p$ is the total number of punctures.
The total degeneracy of the horizon states
    is obtained by summing over all possible values for the
 occupancies $\{ n_j\}$. The pre factor in the right-hand side of above equation comes
from the fact that the spin $j$ can be placed on any of the  $n_j$   sites from the set of all $p
 $ punctures ref\/lecting the fact that the punctures are distinguishable. Formula~(\ref{degen}) can be rewritten as:
\begin{gather*}
 {\cal N} (\{ n^{}_j\}) =  \frac{\left(\sum n_j\right)!}{\prod  n^{}_j !}  \left[ I^{}_0(\{n_j\}) - I_1^{}(\{n_j\})\right],
\end{gather*}
where
\begin{gather*}
I_0^{}   \equiv \frac{1}{k+2}
 \sum^{k+1}_{\ell = 1,2,\dots }
 \prod_j \left[ \frac{\sin\left( \frac{(2j+1)\th^{}_\ell}
{2}\right)}{\sin \frac{\th_\ell}{2}} \right]^{n_j}
= \frac{1}{k+2}  \sum^{k+1}_{\ell = 1,2,\dots }  \prod^{}_j  \left[ \sum_{m=-j}^{j} e^{im\th_\ell^{}}
\right]^{n^{}_j},\nonumber \\
I_1^{}   \equiv  \frac{1}{k+2} \sum^{k+1}_{\ell = 1,2,\dots }
\cos \th_\ell \prod_j \left[ \frac{\sin\left( \frac{(2j+1)\th^{}_\ell}
{2}\right)}{\sin \frac{\th_\ell}{2}} \right]^{n_j}
= \frac{1}{k+2}  \sum^{k+1}_{\ell = 1,2,\dots } \cos \th^{}_\ell  \prod_j^{}   \left[ \sum_{m=-j}^{j} e^{im\th_\ell^{}} \right]^{n^{}_j}.
\end{gather*}
The quantity $I_0^{}-I^{}_1$ simply counts the number of possible $SU(2)$ singlets in the product representations
of spins $j$ with occupancies~$n_j$ on the punctures. Here~$I_0^{}$ counts the numbers of ways the `magnetic'
quantum numbers $m$  can be put on the various punctures so that their sum is zero and~$I_1$ counts the
corresponding number
where their sum is~$+1$ or equivalently, the sum is~$-1$.

The relevant conf\/igurations are  obtained by maximizing  the entropy $S =
 \ln  {\cal N}(\{n_j\})$
  subject to the
constraint that the   horizon area has a f\/ixed large value $  A_{\rm H}
= 8\pi\g \sum n_j \sqrt{j(j+1)} $ in the $\ell_P^{}=1$ units.
This is  done through solving the maximization condition for $n_j$:
\begin{gather}
\delta \ln {\cal N}(\{ n_j\}) - \frac{\l}{8\pi\g  }  \delta A_{\rm H}^{} =0, \label{maxcond}
\end{gather}
 where $\l$ is a Lagrange multiplier. This f\/inally leads to a formula for the horizon entropy. The
    techniques developed in~\cite{kmprl, dkmprd} and~\cite{gm} can be easily extended to perform the computations in
    a~simple and   straight forward manner for large area, $A_{\rm H}\gg 1$. We now outline these calculations.

Using Stirling's  formula for  the factorial of a large number, equation~(\ref{maxcond}) can be solved
for large~$n^{}_j$  to yield:{\samepage
\begin{gather}
 f_j^{} \equiv \frac{n_j^{}}{\sum_j^{} n_j}  = \exp \left[ -\l \sqrt{j(j+1)}
 +  \frac{\delta \ln I}{\delta n^{}_j}\right], \label{fj3}
\end{gather}
where $I \equiv I_0^{}(\{n_j^{} \}) - I_1^{}(\{n_j^{} \})$.}

For large areas, calculation can be done in the large  $k$ $\big(=\frac{A_{\rm H}}{4\g}\big)$
limit where the summations in $I_0$ and $I_1$
can be replaced by integrals: $\th_\ell = \frac{ 2 \pi\ell}{k+2}
\rightarrow x$ and $\frac{1}{k+2} \sum\limits_{\ell=1, 2, \dots }^{k+1} \rightarrow \frac{1}{2\pi} \int^{2\pi}_0 dx$.
Writing these integrals as: $I_0^{} \approx \frac{1}{2\pi} \int dx \exp \left[F(x)\right]$
and  $I_1^{} \approx \frac{1}{2\pi} \int dx \exp \left[F(x)
+\ln \cos x\right]$ with $F(x) \equiv \sum^{}_j n^{}_j \left[ \ln \sin \left( (2j+1) \frac{x}{2}\right)
- \ln \sin\frac{x}{2} \right]$, these   can be readily evaluated by the steepest
descent method. Each of
  $F(x)$ and $F(x) + \ln \cos x$  has  a maximum  at   $x=0$. Evaluating the integrals
  by quadratic f\/luctuations around this maximum  point leads to:
\begin{gather*}
I_0^{}(\{n_j\})  \approx  C  \frac{ \exp\left[F(0)\right] }{\sqrt{ -F''(0)}}  , \qquad
I^{}_1(\{n_j\}) \approx C   \frac{\exp\left[ F(0)\right] }{\sqrt{ -F''(0)  +1}} , \nonumber \\
I (\{n_j\})  \equiv   I_0^{}(\{n_j\})- I^{}_1(\{n_j\}) \approx  \frac{C}{2}  \frac{\exp\left[ F(0)\right] }{ \left[ -F''(0)
\right]^{3/2}_{}} ,
\end{gather*}
where $C$ is just a constant and
\begin{gather*}
 F(0)  =  \sum^{}_j n^{}_j \ln  (2j+1),   \qquad  -F''(0) = \frac{1}{3} \sum_j^{} n_j^{}  j(j+1) =
  \frac{\a A_{\rm H}}{4},
\\
   \a   \equiv  \frac{1}{6\pi \g} \left( \frac{\sum^{}_j f^{}_j \ j(j+1) }{\sum_j^{} f_j  \sqrt{j(j+1)}}\right).
\end{gather*}
 Here we have used the relation between the total number of punctures $p=\sum_j^{}n^{}_j$ and the horizon area:
 $A_{\rm H} = 8\pi\g p \sum^{}_j f^{}_j \sqrt{j(j+1)}$.
  From the results above, we have
\begin{gather*}
\frac{\delta \ln I}{\delta n^{}_j} \approx \ln(2j+1) - \left[ \frac{2j(j+1)}{\a A_{\rm H}}\right],
\end{gather*}
 so that, from equation~(\ref{fj3}),  for the relevant conf\/igurations, the probability distribution for
   the fraction $f_j^{}$  of the puncture sites with spin $j$ can be written as:
 \begin{gather}
 f_j^{} \equiv \frac{n_j^{}}{\sum_j^{} n_j}   \approx  (2j+1)\exp \left[ -\l \sqrt{j(j+1)} -\left( \frac{2j(j+1)}{\a A_{\rm H}}\right)
   \right]. \label{fj4}
\end{gather}
This provides an   ${\cal O} \left(A^{-1}_{\rm H} \right)$ correction to the  formula (\ref{fj}).
 Further, this probability distribution leads to the degeneracy of the horizon states given by:
\begin{gather*}
 {\cal N}^{}_{}(\{n_j\})   \approx  \left(\frac{ 4 }{A_{\rm H}} \right)^{3/2}
 \exp\left[ \left(\frac{\l}{2\pi\g} \right) \frac{A_{\rm H}}{4\ } \right] ,
 \end{gather*}
 whose logarithm  yields the  entropy formula for a large area black hole.

 It is remarkable that this more careful calculation yields  an entropy formula which has exactly
 the same form as  that in equation~(\ref {en3}) obtained
 earlier from the simplif\/ied calculation   where all the punctures were assigned a common low value $r$ of spin.
 As has been pointed out in Section~\ref{III.1},   the coef\/f\/icient~$-3/2$ of logarithmic correction is
  not sensitive to this common value~$r$ of spin used in the calculations there. It is, therefore, no surprise that
  the more careful computation outlined above yields the same value for this coef\/f\/icient.    On the other hand, in the calculations of Section~\ref{III.1}, the coef\/f\/icient of the linear area term does   depend on the common low value~$r$ of the spin assigned to all the sites. In the more accurate calculation above also, we f\/ind
   that this coef\/f\/icient indeed depends on  how spins are distributed
 on the puncture sites. Fixing this coef\/f\/icient by matching with the Bekenstein--Hawking
 area formula yields $\g = \frac{\l}{2\pi}$. An important conclusion that follows from the improved
 calculations   is that, from the distribution~(\ref{fj}) or~(\ref{fj4}) for the maximized entropy,
 we   now have a   reliable estimate of the coef\/f\/icient of the linear area term and consequently that of the
 Barbero--Immirzi parameter. Thus, for the  value $\l \approx 1.72$   obtained by solving~(\ref{fj1}),
 we have  $\g =\frac{\l}{2\pi} \approx 0.27$. This value for~$\g$ was f\/irst reported in~\cite{gm}, where
 the computations were done in the $U(1)$ framework in a way which is equivalent to the calculations above
 with~$I^{}_1$ put to zero, resulting in  the coef\/f\/icient of the logarithmic term   as~$-1/2$ in contrast to its  correct value~$-3/2$.

\subsection[$SU(2)$ versus $U(1)$ formulations]{$\boldsymbol{SU(2)}$ versus $\boldsymbol{U(1)}$ formulations} \label{III.3}

The $U(1)$ and $SU(2)$ Chern--Simons descriptions of horizon   are
some times viewed as counterpoints to each other and entropy calculations, particularly the
logarithmic correction,  in these
two frameworks have been erroneously claimed to yield dif\/ferent results.
These two points of view are in fact quite reconcilable. We have discussed this in the classical context
earlier in Section~\ref{II}:  $U(1)$ formulation is merely  a partially gauge f\/ixed~$SU(2)$ theory. As demonstrated there and  emphasized earlier in~\cite{bkm},
there are additional constraints in the~$U(1)$ description.
  The Isolated Horizon boundary conditions
 which imply the $U(1)$ Chern--Simons equations~(\ref{IHCS}) where the   source $\S^{(1)}_{\th\p}$
   is  one  of the   components  of  $SU(2)$ triplet of solder forms $\S^{(i)}_{\th\p}
 =\g^{-1}_{} e^{ijk}_{} e^i_\th e^k_\p$,  also   imply the constraints (\ref{addcond})
 for the solder forms $\S^{(2)}_{\th \p}$ and $\S^{(3)}_{\th \p}$ orthogonal to the $U(1)$ direction.
  These constraints essentially ref\/lect the $SU(2)$ underpinnings of the classical $U(1)$ formulation.
The quantum theory would have corresponding quantum version of these constraints.
The  horizon properties like associated   entropy
 calculated from the quantum versions of these equivalent theories have to be exactly same.
 This is so as  physics can not change by a gauge f\/ixing in a gauge theory. To see that this
is indeed so,  care needs
 to be  exercised by taking the quantum analogue of the classical conditions (\ref{addcond}) into
account in the calculations  done in the  $U(1)$ framework. These additional conditions play a
crucial role  in  the  computations and when  properly implemented,
  exactly the same formula (\ref{en3})
 for   micro-canonical entropy of large area horizons follows in a straight forward manner.
In the following, we shall  brief\/ly present an outline of this computation \cite{bkm}.

We wish to study the quantum formulation of  the classical  boundary $U(1)$ theory based on
the action (\ref{CSU1}) and equations of motion (\ref{F1}) described in terms of the classical
constraint: $-\frac{k}{2\pi} F^{}_{\th \p}
= \S^{(1)}_{\th\p}$, where $F^{}_{\th \p} =
\partial^{}_{\th} A^{}_{\p}- \partial^{}_\p A^{}_{\th}$ is the f\/ield strength
of the boundary dynamical $ U(1)$ gauge f\/ield $A^{}_{\hat  a}$  $(\hat a = \th, \p)$.
  In the quantum  boundary theory, the f\/ields $(A^{}_\p,  A^{}_\th)$ form
a~mutually conjugate canonical pair with commutation relation $\left[ A_\p (\s_1), A_\th (\s_2)\right] = \frac{2\pi i}{k}
 \delta^{(2)} (\s_1,\s_2)$.
The solder form $\S^{(1)}_{\th\p}$  is not a dynamical  f\/ield in the
boundary theory where it acts merely  as an
{\it external source}. On the other hand, the  bulk quantum theory described by LQG
 is set up in terms of cylindrical functions made up of Wilson line operators of the
bulk $SU(2)$ gauge f\/ields  as the conf\/iguration operators.
The corresponding momentum operators are  the f\/luxes with two-dimensional smearing
$\int_{\mathbb{S}^2} d^2\s  \S^{(i)}_{\th\p}$ so
that their Hamiltonian vector f\/ields map cylindrical functions to cylindrical functions.  The
solder forms $\S^{(i)}_{\th\p}$ have distributional support at the punctures where the
spin networks impinge on the horizon. The  hole micro-states $|\Psi \rangle $ of the quantum theory
are composite states of the boundary theory and those of the bulk: $|\Psi \rangle
\equiv$ $|{\rm boundary}\rangle \otimes |{\rm bulk}\rangle$ where the   operators of the boundary theory
act on the former and those of the bulk on the latter.
 Analogue of the classical IH boundary constraint of~$U(1)$ formulation
mentioned above has to be written in the quantum theory in terms of
  the f\/lux operator  (which is a dynamical operator of the bulk theory)  instead of the solder
form itself. Thus the physical states $|\Psi \rangle$   satisfy the quantum
constraint: $ - \frac{k}{2\pi}\int_{\mathbb{S}^2} d^2\s  F^{}_{\th \p}  |\Psi \rangle
= \int_{\mathbb{S}^2} d^2\s  \S^{(1)}_{\th\p} |\Psi \rangle$,
which relates the   quantum f\/lux through $\mathbb{S}^2$   of horizon in the boundary $U(1)$ gauge theory
 with the quantum f\/lux of the bulk theory.  Next,
the $U(1)$ f\/lux $\int_{\mathbb{S}^2} d^2_{}\s  F^{}_{\th\p}$
in the boundary theory gets contributions, due to Stokes' theorem, from holonomies associated
with  the $p$  punctures as $\sum\limits^p_{i=1}
\oint^{}_{C_{(i)} }d\s^{\hat a}_{} A^{}_{\hat a} $ where ${\hat a}= (\th, \p) $ and
$\{C_{(i)}\}$ are $p$ non-intersecting loops in $\mathbb{S}^2$ such that   $i$th puncture is enclosed
by a~small loop  $C^{}_{(i)}$.  This f\/lux is
equivalently given by the holonomy associated with  a large  closed loop $C$ enclosing all the
punctures: $\oint^{}_C d\s^{\hat a}_{} A^{}_{\hat a} $. Now, on $\mathbb{S}^2$, this large loop $C$ can be continuously shrunk on the other side to
a point without crossing any of the punctures.
This implies that the $U(1)$ f\/lux $\int_{\mathbb{S}^2} d^2_{}\s  F^{}_{\th\p}$   is zero on the physical
states.
Due to the quantum operator constraint  mentioned above, this
 in turn  leads to  the fact that
the bulk f\/lux  operator $\int_{\mathbb{S}^2} d^2_{}\s  \S^{(1)}_{\th\p}$  annihilates the physical
horizon states $ |\Psi \rangle$:
\begin{gather}
 - \frac{k}{2\pi} \int_{\mathbb{S}^2} d^2\s  F^{}_{\th \p}  |\Psi \rangle  =
\int_{\mathbb{S}^2} d^2\s  \S^{(1)}_{\th \p}  |\Psi \rangle  =  0.     \label{quant}
\end{gather}

Similarly, the quantum analogue of the additional classical conditions (\ref{addcond}) have to be written,
instead of the solder forms $\S^{(2)}_{\th\p}$ and  $\S^{(3)}_{\th\p}$,
in terms of the dynamical operators of the bulk theory, namely the f\/lux operators
which act as derivations on the cylindrical functions (functionals of the holonomies).
Consistent with these properties,
the bulk f\/luxes acting on physical states $|{\Psi}\rangle$ of a spherically symmetric  horizon,
in addition to~(\ref{quant}), satisfy the constraints:
\begin{gather}
 \int_{\mathbb{S}^2} d^2_{}\s  \S^{(2)}_{\th\p} |\Psi \rangle  = 0 , \qquad
\int_{\mathbb{S}^2} d^2_{}\s  \S^{(3)}_{\th \p} |\Psi \rangle   =  0.    \label{quantcond}
\end{gather}

As stated earlier,  the properties of the quantum operators  $\S^{(i)}_{\th\p}$ are completely
determined by the bulk theory  where these solder forms, acting on the spin networks,
 are distributional  with support at the punctures   where the network impinges on $\mathbb{S}^2$
 of the horizon.   In particular,
the  bulk f\/lux operators, acting as derivations on spin network states of the bulk theory,
  satisfy a~non-trivial commutation algebra:
$\left[ \int_{\mathbb{S}^2} d^2_{}\s  \S^{(i)}_{\th \p}  ,   \int_{\mathbb{S}^2} d^2_{} \s  \S^{(j)}_{\th \p}
\right] = i \e^{ijk}_{} \int_{\mathbb{S}^2} d^2\s  \S^{(k)}_{\th\p}$.
The quantum   constraints~(\ref{quantcond}) are consistent with this property.
  These  additional conditions  only  ref\/lect the underlying $SU(2)$ invariance
of  the gauge f\/ixed quantum boundary theory  described in terms of~$U(1)$  gauge f\/ields and
are merely the quantum analogues
 of  the constraints~(\ref{addcond}) of the classical  $U(1)$ theory written in terms of the dynamical
operators  of the bulk quantum theory.

Conversely, it is straightforward to check that  the additional conditions (\ref{quantcond}) of the
$U(1)$ formulation  indeed follow directly  from the partial gauge f\/ixing of the {\it quantum}
$SU(2)$ Chern--Simons  boundary theory.
This $SU(2)$ formulation  is  described by a  quantum constraint in terms of the exponentiated form
of the f\/lux operators acting on the physical states
$|\Psi \rangle$ as: ${\cal P} \exp \left( \int_{\mathbb{S}^2} d^2_{}\s
 \S^{(i)}_{\th\p}T^{(i)}_{}\right)|\Psi \rangle = {\cal P} \exp\left( -  \frac{k}{2\pi}\int_{\mathbb{S}^2}
d^2_{}\s  F^{ }_{\th\p}\right) |\Psi \rangle $,  where $T^{(i)}_{}$ is a
basis of $SU(2)$ algebra and $F^{}_{\th\p} \equiv F^{(i)}_{\th \p} T^{(i)}_{}$
is the f\/ield strength of the boundary $SU(2)$ gauge f\/ield $A^{}_{\hat a} \equiv A^{(i)}_{\hat a} T^{(i)}$.
Here the symbol ${\cal P}$ represents {\it surface $($path$)$ ordering} in a specif\/ic way
consistent with the non-Abelian Stokes' theorem \cite{arefeva, sahlmannthiemann}.
This constraint relates the  f\/lux functional of the bulk theory to that
of the boundary gauge theory. The $SU(2)$ gauge transformations act on these f\/lux functionals
as conjugations.  In the boundary Chern--Simons theory, $SU(2)$ quantum gauge f\/ields
$(A^{(i)}_\p,  A^{(i)}_\th)$ are mutually conjugate with their commutation relations as:
$\big[A^{(i)}_\p (\s^{}_1) , A^{(j)}_\th (\s^{}_2) \big]   = \frac{2\pi i}{k} \delta^{ij}_{}
\delta^{(2)}_{}(\s^{}_1, \s^{}_2)$.
Consequently, the boundary gauge f\/luxes $-\frac{k}{2\pi}\int_{\mathbb{S}^2} d^2\s  F^{(i)}_{\th\p}$
do not commute;  in fact,  it is easy to check that these obey $SU(2)$ Lie algebra commutation rules:
\[
\left[\frac{k}{2\pi}\int_{\mathbb{S}^2} d^2\s   F^{(i)}_{\th\p},  \frac{k}{2\pi}\int_{\mathbb{S}^2}
d^2\s  F^{(j)}_{\th\p}\right]
 = -  i\e^{ijk}_{}   \frac{k}{2\pi}\int_{\mathbb{S}^2} d^2\s  F^{(k)}_{\th\p}.
 \] In a similar manner,\looseness=-1
as pointed out earlier, the bulk  f\/lux operators $ \int_{\mathbb{S}^2} d^2\s \S^{(i)}_{\th \p}$ also
satisfy $SU(2)$ Lie algebra commutation rules. Therefore, this introduces
ordering ambiguities  in the def\/inition of the surface
ordered  boundary and bulk f\/lux functionals used here. However, these  ambiguities  can be
f\/ixed by using  the Duf\/lo map  which provides a  quantization map  for functions on Lie algebras
\cite{duflo, sahlmannthiemann}. Now, a non-Abelian generalization \cite{arefeva}
of the Stokes' theorem  allows us to replace the surface ordered boundary  gauge f\/lux functional
depending on the $SU(2)$ f\/ield strength by   a related  path ordered  holonomy functional
of the corresponding  boundary $SU(2)$ gauge connection: ${\cal P} \exp \left( - \frac{k}{2\pi}\int_{\mathbb{S}^2}
d^2\s  F^{}_{\th\p} \right)
= {\cal P} \exp \left( - \frac{k}{2\pi}\oint^{}_C d\s^{\hat a}_{}A^{}_{\hat a}\right) $,  where $C$ is a
contour enclosing all the punctures. Since this
contour~$C$ can be contracted to a point on $\mathbb{S}^2$, this holonomy functional  is simply equal to~$1$
on the physical states so that the quantum f\/luxes of the bulk and boundary theories
acting on the physical states $|\Psi \rangle$ satisfy the
constraint:
\[
{\cal P} \exp \left( \int_{\mathbb{S}^2} d^2_{}\s
 \S^{(i)}_{\th\p}T^{(i)}_{}\right)|\Psi \rangle
  = {\cal P} \exp\left(- \frac{k}{2\pi} \int_{\mathbb{S}^2} d^2_{}\s  F^{(i)}_{\th\p}T^{(i)}_{}\right)
 |\Psi \rangle   = |\Psi\rangle.
 \] The degeneracy of the black hole quantum states may be calculated
 in the boundary $SU(2)$  Chern--Simons theory by counting those states where  the functional of  gauge
 f\/lux operator evaluated over the punctured~$\mathbb{S}^2$ of the horizon
 has eigenvalue~$1$. The result is obtained by counting of the number of ways singlets can be constructed by
 composing the spins~$j_i$  on the punctures in the $SU(2)$  Chern--Simons theory.
 This is exactly how the computations outlined in Sections~\ref{III.1} and~\ref{III.2} have been performed.
  Equivalently, this degeneracy may also be calculated by counting the bulk  spin network states
 on which the bulk f\/lux functional has eigenvalue~$1$. Note that the punctures
 carrying the spins~$j_i$ on the~$\mathbb{S}^2$ of the horizon are common to the boundary and bulk states
 which are connected by  the functional f\/lux constraint. This ensures that the counting done in
 these two ways, in the boundary theory and in the bulk theory, yield the same results.
 Next, the boundary $SU(2)$ quantum Chern--Simons theory can be partially gauge f\/ixed to a~gauge
 theory based on the maximal torus group $T = U(1)$   of~$SU(2)$  through
 appropriate  gauge conditions on the boundary gauge f\/ields such that
the gauge f\/luxes in the two  internal directions orthogonal to this $U(1)$ subgroup
are zero: $\int_{\mathbb{S}^2} d^2_{}\s F^{(2)}_{\th\p}=0$
and $\int_{\mathbb{S}^2} d^2_{}\s F^{(3)}_{\th\p} =0$.
 This Abelian reduction  converts
 the  f\/lux  constraint   of the $SU(2)$ formulation to that   of the $U(1)$ formulation:
\[
\exp\left(\int_{\mathbb{S}^2} d^2_{}\s
 \S^{(1)}_{\th\p}\right)|\Psi \rangle = \exp \left( -\frac{k}{2\pi} \int_{\mathbb{S}^2} d^2_{}\s
 F ^{}_{\th\p}\right) |\Psi \rangle = |\Psi\rangle,
 \]
where now $ F^{}_{\th\p}$ here is the f\/ield strength of the boundary~$U(1)$
gauge f\/ield,  {\it along with the
additional  quantum  conditions for the bulk f\/luxes \eqref{quantcond}}:
 $\int_{\mathbb{S}^2} d^2_{}\s  \S^{(2)}_{\th\p} |\Psi \rangle =0$
and $\int_{\mathbb{S}^2} d^2_{}\s  \S^{(3)}_{\th \p} |\Psi \rangle   =  0$.

Now the horizon entropy  in the $U(1)$ formulation  is obtained by counting, for
 a f\/ixed large area,  the number of ways the spins $j^{}_1,  j^{}_2, \dots , j^{}_p$ can be placed on
 the~$p$ punctures  so that their $U(1)$ projection eigenvalues, the $m$-quantum numbers of
 the diagonal f\/lux operator  $\int_{\mathbb{S}^2}  d^2\s \S^{(1)}_{\th \p}$,  add up to zero,
$m^{}_{\rm tot}\equiv \sum\limits^{p}_{l=1} m^{}_l =0$, to ensure that the physical states $|\Psi \rangle$
satisfy the  $U(1)$ constraint (\ref{quant}).
Notice that these $m^{}_{\rm tot} =0$ conf\/igurations include all such states from the
irreducible representations with spin $j^{}_{\rm tot} = 0, 1,2,3, \dots $ in the tensor
product $\otimes^p_{l=1} (j^{}_l)$.
 Total number of these conf\/igurations  is  counted  exactly by  the f\/irst term of the
 degeneracy formula~(\ref{NP5}) in the large
 $k$ limit ($k\gg 1$) where the periodic Kronecker delta $\bar {\delta}_{m,n}^{}$ becomes
 ordinary   delta $\delta^{}_{m,n}$. Now, if we ignore the
additional constraints (\ref{quantcond}) of the quantum theory,
 from the f\/irst term of~(\ref{NP5}),  we shall get the entropy with the leading area term
 and a sub-leading
 $\ln A_{\rm H}$ correction with coef\/f\/icient $-1/2$ as has been done in several places
\cite{dkmprd,meiss,gm}.  But   this is clearly {\it an over counting }
 of the horizon states as the correct counting would require to exclude those states with
   $ \sum\limits^{p}_{l=1} m^{}_l=0$  which do not respect these additional
constraints. To reiterate, the
 additional quantum constraints (\ref{quantcond}) require that only physical states to be counted
for a non-rotating horizon are
 those which belong to the kernel of the ladder generators $J^{(\pm)} \equiv  \int_{\mathbb{S}^2} d^2\s
\frac{1}{2}\left[
\S^{(2)}_{\th\p} \pm i\S^{(3)}_{\th\p}\right]$
 of the total spin algebra,  besides being in  the kernel of the diagonal  generator $J^{(1)}
\equiv \int_{\mathbb{S}^2} d^2\s \S^{(1)}_{\th\p}$ with  eigenvalues $m^{}_{\rm tot} =0$.
 Now, acting on some of the  vectors in the kernel of the diagonal generator,
 the ladder generators $J^{(\pm)}_{}$  map them to states  with $m$-quantum number,
  $m^{}_{\rm tot} \equiv \sum\limits^p_{l=1}m^{}_l=\pm 1$. These are the states with $m^{}_{\rm tot}=0$ from
  total spin $j^{}_{\rm tot} = 1, 2, 3 ,\dots $
  states   in the tensor product representation $\otimes^p_{l=1} (j^{}_l)$. Thus
  the quantum  constraints (\ref{quantcond}) require that
  such states  {\it should not be included}  in the count. The second and third terms in the formula
  (\ref{NP5}) (in the large $k$ limit) precisely count these extra states. Since the $m^{}_{\rm tot} =+1$
  and $m^{}_{\rm tot} =-1$
  components occur in a non-zero integer spin state in equal numbers,  we have the
   normalization coef\/f\/icient $1/2$ in front of each of these two terms in (\ref{NP5}).
   Thus, even in the gauge f\/ixed   formulation  described in terms of quantum $U(1)$ Chern--Simons
theory with  correctly identif\/ied additional quantum constraints (\ref{quantcond}), a careful counting
  leads to the same formula (\ref{NP5}) as in  the $SU(2)$
   framework;   the asymptotic entropy formula (\ref{en3}) with the coef\/f\/icient $-3/2$ for the leading
   log(area) correction holds in both the formulations. Additionally,  the value of
   Barbero--Immirzi parameter $\g$ f\/ixed through matching of the area term with Bekenstein--Hawking
law is  also the same. These results are not surprising but merely a~ref\/lection of the fact that
   gauge invariance requires that physical quantities  do not change by gauge f\/ixing.

 In fact, like in any other gauge theory, we could further f\/ix the gauge in the $U(1)$ formulation
 so that whole of  the gauge invariance of the boundary theory is now f\/ixed.
 Gauge invariance would imply that counting of the relevant states in this formulation
 should again  yield the same result for black hole entropy as that in the formulation with
 full $SU(2)$ gauge invariance.

\section{Black hole entropy from other perspectives \label{IV}}

Black hole entropy has also  been  calculated in quantum frameworks  other than that provided by LQG.
 These lead to  several   derivations of the asymptotic entropy formula   (\ref{en3})  for a variety of black holes.
  This includes those for many black holes in the String Theory.
This entropy formula appears to hold even for black holes of theories
in dimensions other than four. We shall brief\/ly survey a few of these cases here.

\subsection{Entropy from Cardy formula \label{IV.1}}
Immediately after the   discovery of  $-(3/2) \ln A_{\rm H}$ correction to the Bekenstein--Hawking
area law  obtained from the $SU(2)$ Chern--Simons theory of horizon in  LQG~\cite{kmprl},
Carlip demonstrated that this is  in fact  a generic feature of any conformal f\/ield theory independent
of its  detail structure~\cite{ca}. This important result was derived by a careful calculation of the
logarithmic correction
to the Cardy formula.   The number density of states $\r(\Delta)$ with the eigenvalue $\Delta$ of the
generator~$L^{}_0$ of Virasoro algebra  in a conformal f\/ield theory with central charge $c$, for
large~$\Delta$,  was shown to be:
  \begin{gather}
 \r(\Delta)   \sim  \left( \frac{c }{96 \Delta^3_{}} \right)^{1/4}
 \exp \left( 2\pi  \sqrt{ \frac{c \Delta  }{6} } \right).       \label{carlipdensity}
 \end{gather}
 The exponential term is the Cardy formula \cite{cardy} and the fore-factor
 provides logarithmic correction to it.
Derivation of this result does not require any detail knowledge of the
    partition function of  conformal f\/ield theory; all that goes in to the calculations
  is the generic modular transformation properties of the torus partition function.

  The Carlip formula (\ref{carlipdensity}) is of particular  interest as there are
 strong suggestions that   conformal f\/ield theories do indeed provide a universal
 description of low energy properties of black holes~\cite{bhlms} which is relevant even in
 the   framework of   String Theory.  For the case where black hole horizon properties are described by
 a single conformal theory, the argument of the exponential in~(\ref{carlipdensity}) can be identif\/ied
 with the Bekenstein--Hawking entropy, $S^{}_{\rm BH} =2\pi \sqrt{c\Delta/6}$. This
   is the case with many black holes    in the String Theory.  Thus, for
 such a model, the Carlip  formula    readily yields horizon entropy as:
 \begin{gather}
 S^{}_{\rm H}  = \ln \r(\Delta)  =  S^{}_{\rm BH}  -  \frac{3}{2}  \ln S^{}_{\rm BH}  +  \ln c  +  \cdots
\label{carlipentropy}
 \end{gather}
  with its f\/irst two terms same as in the LQG entropy formula~(\ref{en3}). Carlip has applied this result   to
  analyse  several cases
   which include  the BTZ black hole in $2+1$ dimensions and string theoretic counting of D-brane states for BPS
 black holes~\cite{svh}. With this, presence of logarithmic correction with the def\/inite coef\/f\/icient~$-3/2$
  for many black holes in three, four and higher dimensions   has been    established.
 Carlip has made an eloquent case for the {\it universal}
nature of this logarithmic correction.

 In  another derivation of  black hole  entropy from the conformal f\/ield theory perspective,
 instead of the corrected Cardy formula,
 Rademacher's exact convergent expansion  for the Fourier coef\/f\/icients of a modular form of a given weight
 has been used in \cite{bs}. This analysis also shows that, for large holes, the leading logarithmic correction
 to the entropy has the  universal coef\/f\/icient~$-3/2$,  again in conformity with     the   LQG  formula~(\ref{en3}).

  \subsection{Entropy of   BTZ black hole in the Euclidean path integral approach} \label{IV.2}

   There are alternative  methods, besides
  those described above, to study the entropy of BTZ  hole. In fact, it is
  possible to derive  an {\it exact}  expression for the partition function of     Euclidean
  BTZ black hole   in the path integral approach~\cite{gks}.
  Entropy   for a  large area     Lorentzian BTZ   hole is extracted from this
  after a proper analytic continuation.

  We start by writing three-dimensional Euclidean gravity   with a negative cosmological
  constant in the f\/irst order formulation (with triads $e$ and spin connection $\w$) in terms of
  two $SU(2)$ Chern--Simons theories \cite{atwc}:
  \begin{gather}
  I^{}_{\rm grav}  =  k I^{}_{\rm CS}[A]  -   k I^{}_{\rm CS}[{\bar A}] , \label{CSBTZ}
  \end{gather}
  where  $I^{}_{\rm CS}$ represents the Chern--Simons action for complex gauge f\/ields
  $A= \left( i  \ell^{-1}_{}e^i_{} +\w^i  \right) T^i$ and ${\bar A}=
  \left( i  \ell^{-1}_{}e^i_{} -\w^i  \right) T^i$ with $T^i \equiv i\s^i_{}/2$ as the  generators of
  the Lie algebra of $SU(2)$ and coupling $k =\ell/ (4G)$ for negative cosmological constant
  $\Lambda  =- 1/{\ell}^2_{}$. The gauge group of this theory is $SL(2, \mathbb{C})$. Corresponding gauge
  group for Minkowski gravity with negative cosmological constant is $SO(2,\mathbb{R}) \otimes SO(2,\mathbb{R})$
  with coupling $k=-\ell/(4G)$.  The Lorentzian results are obtained from Euclidean theory
  after completing the computations of various quantities of interest by an analytic
  continuation $G \rightarrow -G$.

Taking time and  angular momentum to be pure imaginary as $t=i\t_{\rm E}$ and  $J = -i J_{\rm E}$ and consequently
the inner horizon radius $r^{}_- = -i |r^{}_{{\rm E}-}|$, we obtain the Euclidean continuation
of  BTZ black hole  which
   has the topology of a solid torus~\cite{ct}. On a solid torus,
  of the two non-trivial cycles of the boundary 2-torus, one becomes contractible while other
  is non-contractible. We are
  interested in the   path integral for the Chern--Simons theory (\ref{CSBTZ}) on a solid
  torus with a~boundary modular parameter $\t = \t^{}_1 +i\t^{}_2$. This is evaluated by keeping
  the trace of gauge f\/ield holonomy along the non-contractible cycle on  the {\it boundary}  $2$-torus  f\/ixed.
  This  holonomy  is a~function of  the  outer (event) and inner   horizon radii $r^{}_+$ and $r^{}_{-}$
  which are given in terms of the  mass parameter $M$ and angular momentum $J$ of the  hole
  and hence, this keeps these quantities f\/ixed. The
     quantum f\/luctuations are introduced  through
  Wilson lines with spin~$n/2$   along the non-contractible cycle
  {\it inside} the solid torus. These create  defect angles (characterized by the spin $n/2$)
  at the horizon which are  not kept f\/ixed and all possible   values are included in the path
integral by   summing over various  values of the spin $n/2$ (with $n\le k $).
    The states corresponding to these   closed Wilson lines are  given by~\cite{lab}:
\[
 \psi^{}_n(u, \t)  =  \exp\left( \frac{\pi k u^2_{}}{4\t^{}_2}\right)   \chi^{}_n (u, \t),
 \qquad n=0,1,2, \dots , k ,
\]
  where $u=   -\t -i\left(r^{}_+ + i|r^{}_{-}|\right)/\ell $ characterizes the boundary value of the gauge
  connection  and~$\chi^{}_n$ are the Weyl--Kac characters for af\/f\/ine $SU(2)$ which are given in terms
  of the Theta functions~as:
\[
\chi^{}_n (u, \t)  = \frac{ \Theta^{(k+2)}_{n+1} (u, \t, 0) - \Theta^{(k+2)}_{-n-1}(u,\t,0)}
  {\Theta^{(2)}_1(u,\t,0) -\Theta^{(2)}_{-1}(u,\t,0)}
\]
  with
\[
 \Theta^{(k)}_n(u,\t,z)  =  \exp \left(-2\pi ik z\right)  \sum^{}_{s \in {\mathbb Z}} \exp\left\{ 2\pi i k
  \left[ \left( s+ \frac{n}{2k}\right)^2 \t +\left(s + \frac{n}{2k} \right)u \right] \right\}.
\]
  Finally, the Euclidean black hole partition function is given by \cite{gks}:
\begin{gather}
 Z_{\rm E}  =  \int d\m(\t, {\bar \t})  \left| \sum^{k}_{n=0} \left(\psi^{}_n(0,\t)\right)^*_{}
  \psi^{}_n (u, \t) \right|^2_{}. \label{ZE}
\end{gather}
  Here   the integrand is
invariant under the $2$-torus modular transformations ${\cal S}:\t \rightarrow -1/\t$,
$ u\rightarrow u/\t$ and ${\cal T}: \t \rightarrow \t+1$.  The integration is done
with the modular invariant measure: $d\m (\t, {\bar \t}) = (d\t d{\bar \t})/\t^2_2$
where $\t^{}_2 = {\rm Im}\, \t$.

The result (\ref{ZE}) is an {\it exact} expression for the partition function of a Euclidean black hole.
We may now evaluate it for large horizon radius $r^{}_+$ ( $r^{}_+ > \ell > 4G$) by the saddle point method.
The saddle point of the integrand is at a large value of ${\rm Im}\,\t$ given by  $\t^{}_2 =r^{}_+ /\ell$ for $|r^{}_{-}| \ll  r^{}_+$.
The computation is done for a positive coupling constant $k= \ell/(4G)$ and   in the end, we go over to the
Lorentzian black hole  through the analytic continuation $G\rightarrow -G$. After this analytic continuation,
it can be shown that  spin $n=0$ term  dominates in the sum in~(\ref{ZE}). Finally,
for large horizon length
$r^{}_+$ ($r^{}_+ \gg \ell$ with $r^{}_{-} \ll  r^{}_+$), this procedure leads to the  Lorentzian formula~\cite{gks}:
\[
 Z_{\rm L}  \sim  \frac{\ell^2_{}}{r^2_+}   \sqrt{ \frac{8r^{}_+ G}{\pi \ell^2_{}} }
  \exp\left( \frac{2\pi r^{}_+}{4G}\right),
\]
whose logarithm yields  an   asymptotic formula for  the entropy of  Lorentzian BTZ hole:
\begin{gather}
 S^{}_{\rm BTZ}  =  \frac{ 2\pi r^{}_+}{4G} -  \frac{3}{2}  \ln \left( \frac{ 2\pi r^{}_+}{4G}\right)
 +  \ln k  + \cdots \label{SBTZ}
\end{gather}
 with $k= \ell/(4G)$. The leading Bekenstein--Hawking term  was  already obtained  earlier from other Euclidean calculations~\cite{cskg}. The new computation reviewed here   provided  the sub-leading logarithmic correction
with coef\/f\/icient~$-3/2$ in
agreement with  the results obtained for the BTZ   hole from the corrected
asymptotic Cardy formula~(\ref{carlipdensity})  and also
 with the LQG  result~(\ref{en3})   for the holes in four dimensional  theory.

  As pointed out above, the asymptotic formula  (\ref{SBTZ}) holds for large $r^{}_+$ ($r^{}_+ \gg \ell\gg  4G$).
  However, for smaller $r^{}_+$,
   dif\/ferent results   hold~\cite{GKS}: (i) For $r^{}_+ \sim \ell$, where the saddle point of
   the integrand in (\ref{ZE}) occurs at $\t^{}_2 \sim 1$, the entropy
   is given by $S^{}_{\rm BTZ}= \frac{2\pi r^{}_+}{4G}  - \ln \left(\frac{ 2\pi r^{}_+}{4G}\right) + \cdots $ This is also the entropy
   associated with the cosmological horizon of three dimensional Lorentzian de Sitter space.
   (ii) On the other hand, for $r^{}_+ \ll  \ell$,  we have $S^{}_{\rm BTZ}= \frac{ 2\pi \ell^2_{}}{4r^{}_+ G}
    + \frac{3}{2}  \ln \left(\frac{r^{}_+}{\ell}\right) + \cdots $ which represents the entropy of   AdS gas.

\subsection{Entropy of a highly excited string} \label{IV.3}

  It is  more than two decades now since
 't~Hooft suggested a complementarity between black holes and strings~\cite{thooft}: it may be
 possible to provide a black hole interpretation of strings and conversely, black holes may
 have a string representation. Susskind's idea \cite{suss} that the micro-states of a
Schwarzschild black hole
 could be described by the states of a highly excited string at the Hagedron temperature may be
 viewed as a ref\/lection of this complementarity with more evidence provided in~\cite{hrs}
 and others. This correspondence principle    for the two spectra
   can be understood as follows: As the string coupling $g^{}_{\rm str}
\equiv \left(\ell_{\rm P}/ \ell_{\rm S} \right)^{(d-2)}$
 in $d$ dimensions ($\ell_{\rm P}$ is the Planck length and $\ell_{\rm S}$ is the string scale) increases,  the
 Compton wave-length of a high mass and low angular momentum state of the string
 shrinks to a size smaller than its Schwarzschild radius and it becomes a black hole.
Conversely, as the coupling is reduced, the hole becomes smaller  and at some
stage it is smaller than the string size.
 The metric near horizon loses it meaning and instead of the hole, we have an object which is
 better described as a string state. At some in between stage,   when string and hole sizes
 are equal, either description is possible implying a one-to-one correspondence between
the two spectra~\cite{suss}.
Using the fact that near horizon geometry of a Schwarzschild black hole is a Rindler space,
Susskind has suggested that square root of the oscillator number $\sqrt N$ of the highly excited
string should be identif\/ied with the Rindler energy $E_{\rm R}$ instead of the ADM mass of the hole.
Rindler energy and ADM mass are related by a huge red shift between the stretched horizon
and asymptotic inf\/inity.
 For the Schwarzschild black hole in $d$  dimensions ($d \ge 4$), Rindler energy
is linearly related to the horizon  area: $A_{\rm H} = 8\pi G E_{\rm R}$
 and hence the Bekenstein--Hawking entropy is $S_{\rm BH} = 2\pi E_{\rm R}$.

 To push  this correspondence further, we may calculate the density of states $\r(N)$ with high
 oscillator number~$N$
 in a string theory.  For an open bosonic string moving in $d$ dimensions,
the partition function as a function of
 a complex parameter $\t$ is given by~\cite{rkprd}:
 \begin{gather*}
 Z(\t) =   \left( \frac{1}{-i\t} \right)^{(d-2)/2}  e^{-2\pi i \t a}
\,{\rm tr}\, \exp \left[ 2\pi i \t {\cal N} \right]  ,
 \end{gather*}
 where $c= 24 a = (d-2)$ is the central charge   and ${\cal N}$ is the occupation number operator
 with its eigenvalues represented by~$N$. It
 is given in terms of the oscillator number operators ${\cal N}^{}_m$ as $ {\cal N}
 = \sum\limits^{\infty}_{m=1} m  {\cal N}^{}_m $.  Here ${\cal N}^{}_m =
  \sum\limits^{d-2}_{i=1}  a^{i\dagger}_m a^i_m$
  with $a^{i\dagger}_m$ and $a^i_m$, with  $i = 1,2,\dots ,(d-2)$,   as the standard
oscillator creation and destruction operators
  associated with the transverse $(d-2)$ dimensions in the light-cone gauge.  The eigenvalues
  of the oscillator number
  operator  ${\cal N}^{}_m$ are $0, 1,2,3, \dots $.  The expression for partition function above
  has been  derived   with  care   by   taking in to  account the contribution of  zero modes
   which has resulted in the fore-factor $\left( -i \t \right)^{-(d-2)/2}$. The trace over
the oscillator states is done using number theory techniques involving the number of
partitions of $N$ in terms of positive integers and modular transformation properties
of the partition function.

  We may alternatively  write the partition function  in terms of the level density $\r(N)$
  for the states with eigenvalue $N$ of the number operator ${\cal N}$ as:
 $
  Z(\t)  = \sum\limits^{\infty}_{N=0} \r(N) e^{2\pi i (N-a) \t}_{}
  $ which  can be inverted to write a formula for  $\r(N)$ in terms of the partition function.
  After a modular transformation $\t \rightarrow - 1/\t$, this is then evaluated by the saddle point method
  for large $N$ to   yield  the result~\cite{rkprd}:
\[
 \r(N)  \sim  C    \frac{a \exp \left( 4\pi \sqrt{aN} \right)} {\left(aN\right)^{3/4 }_{} },
\]
where $C$ is an $N$ independent constant.   This equation is only a special case of the   formula~(\ref{carlipdensity}) obtained by Carlip for  a general conformal f\/ield theory.

Now, as suggested by Susskind, we identify  the Rindler energy as $E_{\rm R} =2 \sqrt{aN}$. This
leads to an  entropy formula for
 this highly excited string moving in  $d$ dimensions, given by the logarithm of the  level density $\r(N)$, as:
\begin{gather*}
 S^{}_{\rm str}  =  2\pi E_{\rm R}  -  \frac{3}{2}  \ln E_{\rm R}   +  \ln a  +  \cdots.  
\end{gather*}

Though the asymptotic formula for level density
   above was calculated for an open bosonic string, it is valid in general for any string theory.
Again the logarithmic correction with coef\/f\/icient~$-3/2$  matches with the LQG  entropy formula~(\ref{en3}).
This result may be interpreted   as an additional evidence for the {\it excited string $\leftrightarrow$  black hole} correspondence.

  Similar to  the  string calculation above, a more general study of the asymptotic density of states
  for  open $p$-branes has been done by Kalyana Rama in~\cite{rama}.
  Careful  inclusion of the contributions
  of  zero modes here leads to a logarithmic correction to the entropy with coef\/f\/icient~$ -(p+2)/(2p)$
  which agrees with string result~$-3/2$ for $p=1$.

\looseness=1
 We have surveyed some of the alternative derivations of the asymptotic entropy
 for\-mu\-la~(\ref{en3}).  There  are still  several  others \cite{ggs, davidson},
particularly with the same leading  logarithmic correction as in~(\ref{en3}),
that have appeared over the years.  Of   these the  most recent one is by
Davidson where
a discrete holographic  shell model is proposed for
  a spherical black hole~\cite{davidson}.  Instead of putting them on punctures
or small Planck scale patches on the  horizon, the  degrees of freedom are distributed
holographically in the entire black hole interior
in concentric spherical shells of light sheet unit intervals. The   number of
distinguishable conf\/igurations is given by the Catalan series, which readily  leads
to an asymptotic entropy formula with a logarithmic correction with
coef\/f\/icient~$-3/2$. As pointed out in~\cite{davidson}, it is of interest to note that
Catalan numbers are directly
related to a standard {\it stack data   structure} for storage of information in computer science.
In this context, we may also    recall equation~(\ref{NP1/2})  from Section~\ref{III.1}, which represents the fact that
   $SU(2)$ singlets in the composite representation made of spin~$1/2$ assignments on
all   the~$p$ punctures on~$\mathbb{S}^2$ of horizon are precisely counted by the
Catalan number $C^{}_n = \frac{(2n)!}{(n+1)! n!}$ for $p=2n$.

\subsection{Universality of the logarithmic correction} \label{IV.4}

\looseness=1
 As we have listed above, the leading
logarithmic correction, $-3/2\ln A_{\rm H}$,  to the
Bekenstein--Hawking area law  found f\/irst in   LQG
appears to obtain from a variety of other  perspectives, conceptually distinct
from the LQG framework.
The same logarithmic correction has also been
derived for black holes in theories in dimensions other than four.
For example, BTZ black
holes of three dimensional gravity do exhibit this property.
It is  remarkable, though appa\-rently  mysterious, that
such diverse approaches should lead to the same result.
 For all those models where ultimately
the black hole properties are represented by conformal f\/ield theories, Carlip's work
has demonstrated that
such a  correction     is generic.  This holds for many   black holes  in the string theory.
   Besides this logarithmic term,
conformal f\/ield theory   re\-sult~(\ref{carlipentropy}) also has  a~$\ln c$ correction
depending on the central charge.
Power law dependence of $c$   on the area, would change the coef\/f\/icient~$-3/2$
of the $\ln A_{\rm H}$ term.
Thus for  those theories where central charge   is independent of area,
the LQG asymptotic  formula (\ref{en3}) holds. All these facts
suggest a strong  case for the universal character of the logarithmic
area cor\-rection.

\section{Recent developments \label{V}}

\looseness=1
In last few years, there has been a resurgent interest in the  $SU(2)$  gauge theoretic description
of Isolated Horizons within  LQG framework \cite{per, agui, liv2, sahl, kr}.
The $SU(2)$ Chern--Simons theory as a description  of the horizon  has been re-emphasized in \cite{per}
which has been   followed by further work in \cite{ agui, liv2,   sahl} and others.
This has been presented as an alternative to the $U(1)$ Chern--Simons formulation. However, as
we have seen above these
two formulations provide equivalent descriptions with exactly same consequences.
  A direct representation of the black hole states in terms of $SU(2)$ intertwining operators
has also been explored in~\cite{ kr}. The Hilbert space of these operators is the same as those of
the $SU(2)$ Chern--Simons states in the limit of large
 coupling~$k$ ($k\gg 1$).

\looseness=1
Some of these recent papers \cite{per, agui, liv2}  have recalculated   horizon entropy  in the
$SU(2)$ Chern--Simons framework  and
reconf\/irmed the form of   asymptotic entropy formula (\ref{en3}), particularly with the $-(3/2 )\ln
A_{\rm H}$ correction.  For example,
calculations in f\/irst two papers in~\cite{agui} start  with the standard  integral representation in terms
of $SU(2)$ characters for the  degeneracy of $SU(2)$ singlet states in the composite
representation $\otimes^{p}_{l=1} (j^{}_l)$. This is just an
integral representation of the degeneracy formula (\ref{NP3})  (or equivalently
(\ref{NP4}) or (\ref{NP5})) in the limit of large $k$ $\big(=\frac{A_{\rm H}}{4\g}\gg 1\big)$ where these
$SU(2)_k$ formulae can be approximated by  those for ordinary $SU(2)$. The various spin values $j$
are distributed over the punctures with varying occupancy numbers $n^{}_j$.
The degeneracy of  black hole
states is to be obtained by restricting to conf\/igurations which yield area values
in the interval $[A_{\rm H} -\e , A_{\rm H} +\e]$ with
$A_{\rm H}   = 8\pi \g   \sum\limits^p_{l=1} \sqrt{j^{}_l (j^{}_l +1)}$
for a~reasonable choice of $\e$. To solve various
relevant  combinatorial constraints, powerful number theory techniques
developed earlier~\cite{abv} have
been used  to obtain    generating functions from which the
  degeneracy of    black hole states   has been  extracted through
the  Laplace transform method. This then reproduces, for large area,  the
logarithmic correction with the
def\/inite coef\/f\/icient~$-3/2$ as found earlier in~\cite{kmprl} and reviewed here in
Section~\ref{III.1}. Besides this,
 the formula (\ref{fj1}), earlier found in~\cite{meiss,gm}, is also obtained. Consequently
 the improved value of the Barbero--Immirzi parameter
 mentioned in Section~\ref{III.2} and  f\/irst found in~\cite{gm},  is recovered.
 These calculations have been done using  ordinary  $SU(2)$ counting rules, instead of the full
 $SU(2)_k$ formulae. These are adequate  to yield the leading linear area  and
 logarithmic terms of the asymptotic  formula~(\ref{en3}). An important feature of these
 computations is that, when extended to include ef\/fects depending on the  smaller values
 of~$k$, these may also provide a method to study the properties of small  black holes
  where such ef\/fects would be  important.

Lastly,  an   ef\/fective bulk gravity action of the form $f(R)$ that reproduces  the
asymptotic LQG  black hole entropy with
its logarithmic   correction through the Wald procedure
  has also been derived  recently  in \cite{cm}.

\section{Concluding remarks} \label{VI}

\looseness=-1
We have here surveyed how the Chern--Simons theoretical description of   horizon degrees
of freedom emerges in the LQG. This leads to two formulations of the theory with
 $SU(2)$ and $U(1)$ gauge invariances. It has been demonstrated   that  these provide equivalent
 descriptions; the latter being
only a gauge f\/ixed version of the former. A framework developed
more than a decade ago, that relates the $SU(2)$ Chern--Simons theory to gauged $SU(2)_k$ Wess--Zumino
conformal f\/ield theory,  to compute the horizon entropy has been presented. We have also discussed
  the calculations   in the equivalent quantum $U(1)$ formulation.  When carefully identif\/ied
 additional quantum constraints (\ref{quantcond})  are properly implemented,
this formulation also yields the same results as the earlier $SU(2)$ theory.
This is in conformity with fact that gauge f\/ixing does not change the physical properties.
Besides the Bekenstein--Hawking area term, the micro-canonical entropy possesses  a
leading correction as logarithm of the horizon area with coef\/f\/icient $-3/2$ for large horizons. This
is followed by further corrections which are a constant and terms con\-taining inverse
powers of area. These corrections have their origin in the non-perturbative quantum
f\/luctuations of geometry in contrast to those that come from quantum matter f\/luctuations.
The logarithmic correction for black holes in four dimensional gravity
 has since been obtained  in other frameworks too. These results  appear
to be valid for black holes in other than four dimensions as well. We have
outlined  some of these developments here. The logarithmic correction to the
Bekenstein--Hawking entropy may possibly have a universal character.

Derivation of    leading terms of the   asymptotic entropy formula (\ref{en3})
from the Chern--Simons theory of horizon    does not
require the full force of  $SU(2)_k$ conformal f\/ield theory; these terms emerge
by using the counting rules of ordinary $SU(2)$ which corresponds to the large $k $
limit ($k\gg 1$) of   conformal theory.  However, the terms beyond   logarithmic correction
are sensitive to the smaller  values of~$k$. Further, the general
framework explored through the boundary  conformal theory \cite{kmplb, kmprl}
discussed here needs to be exploited to unravel the properties
of small black holes.  Ef\/fect of the smaller values of $k$ $\big(=\frac{A_{\rm H}}{4\g}\big)$ would be pronounced for such holes
and hence $SU(2)_k$ conformal f\/ield theory   will play important role here.

\looseness=-1
In the survey here,  Isolated  Horizons  have been studied in the f\/ixed
area ensemble where   micro-canonical entropy emerges from the micro-states
associated with   quantum geometry degrees of freedom of horizon.   We have
not discussed the entropy of  radiant black holes  in thermal equilibrium with
their radiation bath.  These are   described by a canonical ensemble with f\/ixed
area and energy. The canonical entropy of such a hole results from both
the quantum geometry f\/luctuations of  horizon as well as the thermal f\/luctuations.
While  the  counting of quantum geometry micro-states  leads to a negative
correction  to  the  Bekenstein--Hawking law,     the  thermal f\/luctuations   due to
exchange of   heat between the hole and its surroundings,   which increase
the uncertainty of  the horizon area, would lead to a positive correction.
We shall now close our discussion with  a few remarks about some  recent results
about canonical entropy of these   holes.   A general analysis  to study radiant spherical holes
has been set up in \cite{dmb, gourmed, acpm}. Canonical entropy of  such
 holes has been calculated \cite{acpm} to be given by the micro-canonical
entropy $S^{}_{\rm micro}(A_{\rm H})$  emerging from quantum geometry degrees
of freedom with an extra  correction due to the thermal f\/luctuations as:
$ - \frac{1}{2} \ln \D (A_{\rm H})$ where $\D(A_{\rm H})$ is given in terms of
the mass function $M(A_{\rm H})$  and micro-canonical entropy by
$ M'(A_{\rm H}) S'^2_{\rm micro}(A_{\rm H}) \D (A_{\rm H})  = a \left[ M''(A_{\rm H})
S'^{}_{\rm micro}(A_{\rm H}) - M'(A_{\rm H}) S''^{}_{\rm micro}(A_{\rm H})\right]$. Here prime
denotes derivative with respect to the argument and $a$ is a positive constant.
Also the heat capa\-ci\-ty is  $C= \left[S'^{}_{\rm micro} (A_{\rm H}) \right]^2_{}
/ \D(A_{\rm H})$. This provides a universal criterion for thermal stability of
holes~\cite{acpm}: positivity of $\D(A_{\rm H})$  or equivalently  the
integrated condition   $M(A_{\rm H}) > S^{}_{\rm micro}(A_{\rm H})$ ensures thermal stability.
Clearly for the Schwarzschild black hole where $M(A_{\rm H}) \sim \sqrt{ A_{\rm H}}$,
 we have $\D <0$  ref\/lecting the usual thermal instability. On the other hand, for AdS
Schwarzschild hole,  the mass-area relation is
$M(A_{\rm H}) = \frac{1}{2} \sqrt{ \frac{A_{\rm H}}{2\pi}} \left( 1+
\frac{A_{\rm H}}{4\pi \ell^2_{}}\right)$ with $-1/\ell^2_{} $ as the cosmological constant.
 Here, for small areas, $A_{\rm H}
<4\pi \ell^2_{}$, where   $M(A_{\rm H})$ has the ordinary Schwarzschild like  behaviour
$ M(A_{\rm H}) \sim {\sqrt A_{\rm H}}$,
we have thermal instability.  For large areas, $A_{\rm H} > 4\pi \ell^2_{}$,
where $M(A_{\rm H}) \sim A^{3/2}_{\rm H}$,  we have $\D  >0$ ref\/lecting thermal stability.
Since for this case  $\D (A_{\rm H}) \sim A^{-1}_{\rm H}$, the canonical entropy for a large area stable
AdS Schwarzschild  black hole is given by $S^{}_{\rm can} = S^{}_{\rm micro}(A_{\rm H})
+ \frac{1}{2} \ln A_{\rm H}  + \cdots = A_{\rm H} /( 4\ell^2_P) -\ln \left[A_{\rm H}/(4\ell^2_P)\right]
+\cdots $.  This result, in fact, holds for any smooth mass function $M(A_{\rm H})$. Now,
in  between the small and large area regions,
the point  $\D=0$  (heat capacity $C$ diverging) occurs  at a critical value of the
horizon area $A^{}_c$ where the mass function
equals the micro-canonical entropy in appropriate  units, $M(A^{}_c) = S^{}_{\rm micro}(A^{}_c)$.
This is the point of Hawking-Page  transition. Thus a~characterization of  this phase
transition without any reference to the classical metric has been obtained.

We close our survey with one last remark. The presence of the logarithmic correction
in the asymptotic horizon entropy
may have consequences in a variety of phenomenon in the theory of gravity. In particular,
these may show up in cosmology.  For example, its implications for entropic inf\/lation have been
explored  recently in \cite{efs}.

\newpage

\subsection*{Acknowledgements}

The author gratefully acknowledges collaborations with T.R.~Govindarajan, P.~Majumdar,
S.~Kalyana Rama and V.~Suneeta which have lead to  many of the results surveyed here.
Thanks are also due to Ghanashyam Date for his useful comments.
The support of the Department of Science and Technology,
Government of India, through  a J.C.~Bose National Fellowship is gratefully acknowledged.

\pdfbookmark[1]{References}{ref}
\LastPageEnding


\begin{thebibliography}{99}
\footnotesize\itemsep=0pt

\bibitem{bh}
Bekenstein J.,
Black holes and entropy,
\href{http://dx.doi.org/10.1103/PhysRevD.7.2333}{\textit{Phys. Rev.~D}}  \textbf{7} (1973), 2333--2346.\\
Bekenstein J.,
Generalized second law of thermodynamics in black hole physics,
\href{http://dx.doi.org/0.1103/PhysRevD.9.3292}{\textit{Phys. Rev.~D}} \textbf{9} (1974), 3292--3300.\\
Bardeen J.M., Carter B.,  Hawking S.W.,
The four laws of black hole mechanics,
\href{http://dx.doi.org/10.1007/BF01645742}{\textit{Comm. Math. Phys.}}  \textbf{31} (1973), 161--170.\\
Hawking S.W.,
Particle creation by black holes,
\href{http://dx.doi.org/10.1007/BF02345020}{\textit{Comm. Math. Phys.}}  \textbf{43} (1975), 199--220. \\
Page D.,
Particle emission rates from a black hole: massless particles from an uncharged, non-rotating hole,
\href{http://dx.doi.org/10.1103/PhysRevD.13.198}{\textit{Phys. Rev.~D}} \textbf{13} (1976), 198--206.\\
Unruh W.G.,
Notes on black-hole evaporation,
\href{http://dx.doi.org/10.1103/PhysRevD.14.870}{\textit{Phys. Rev.~D}}  \textbf{14} (1976), 870--892.

\bibitem{ash1}
Ashtekar A.,  Beetle C., Dreyer O.,  Fairhurst S., Krishnan B.,  Lewandowski J.,    Wi\'sniewski J.,
Generic  isolated horizons and their applications,
\href{http://dx.doi.org/10.1103/PhysRevLett.85.3564}{\textit{Phys. Rev. Lett.}}  \textbf{85} (2000), 3564--3567, \href{http://arxiv.org/abs/gr-qc/0006006}{gr-qc/0006006}.\\
Ashtekar A.,   Beetle C.,   Fairhurst S.,
Mechanics of isolated horizons,
\href{http://dx.doi.org/10.1088/0264-9381/17/2/301}{{\it Classical Quantum Gravity}} {\bf 17} (2000), 253--298, \href{http://arxiv.org/abs/gr-qc/9907068}{gr-qc/9907068}.\\
 Ashtekar A.,  Fairhurst S.,   Krishnan B.,
 Isolated horizons: Hamiltonian evolution and the f\/irst law,
\href{http://dx.doi.org/10.1103/PhysRevD.62.104025}{{\it Phys. Rev.~D}} {\bf 62} (2000), 104025, 29~pages, \href{http://arxiv.org/abs/gr-qc/0005083}{gr-qc/0005083}.

 \bibitem{rkpm}
 Kaul R.K.,  Majumdar P.,
 Schwarzschild horizon dynamics and SU(2) Chern--Simons theory,
\href{http://dx.doi.org/10.1103/PhysRevD.83.024038}{{\it Phys. Rev.~D}} {\bf 83} (2011), 024038, 10~pages, \href{http://arxiv.org/abs/1004.5487}{arXiv:1004.5487}.

\bibitem{th}
Birmingham D.,  Blau M.,  Rakowski M.,  Thompson G.,
Topological f\/ield theory,
\href{http://dx.doi.org/10.1016/0370-1573(91)90117-5}{{\it Phys. Rep.}} {\bf 209} (1991), 129--340.

\bibitem{rk}
Kaul R.K.,  Govindarajan T.R.,  Ramadevi P.,
 Schwarz type topological quantum f\/ield theories,
 in  Encyclopedia of Mathematical Physics,
  Elsevier, Amsterdam, 2006, 494--503, \href{http://arxiv.org/abs/hep-th/0504100}{hep-th/0504100}.

 \bibitem{rovbk}  Rovelli C.,
 Quantum Gravity, \textit{Cambridge University  Monographs on Mathematical Physics},
\href{http://dx.doi.org/10.1017/CBO9780511755804}{Cambridge University Press}, Cambridge, 2004.  \\
 Ashtekar A.,  Lewandowski J.,
 Background independent quantum gravity: a status report,
\href{http://dx.doi.org/10.1088/0264-9381/21/15/R01}{{\it Classical Quantum Gravity}} {\bf 21} (2004), R53--R152,
\href{http://arxiv.org/abs/gr-qc/0404018}{gr-qc/0404018}. \\
Theimann T.,
Modern canonical quantum general relativity,
{\it Cambridge Monographs on Mathematical Physics},
\href{http://dx.doi.org/10.1017/CBO9780511755682}{Cambridge University Press},  Cambridge, 2007. \\
Sahlmann H.,
Loop quantum gravity -- a short review,
\href{http://arxiv.org/abs/1001.4188}{arXiv:1001.4188}.

\bibitem{bkm}
Basu R., Kaul R.K.,  Majumdar P.,
 Entropy of isolated horizons revisited,
\href{http://dx.doi.org/10.1103/PhysRevD.82.024007}{{\it  Phys. Rev.~D}} {\bf 82}  (2010), 024007, 5~pages,
\href{http://arxiv.org/abs/0907.0846}{arXiv:0907.0846}.

\bibitem{per}
Engle J., Noui K., Perez A.,
Black hole entropy and $SU(2)$ Chern--Simons theory,
\href{http://dx.doi.org/10.1103/PhysRevLett.105.031302}{{\it Phys. Rev. Lett.}} {\bf 105} (2010), 031302, 4~pages,
\href{http://arxiv.org/abs/0905.3168}{arXiv:0905.3168}. \\
 Engle J.,   Noui K.,  Perez A.,  Pranzetti D.,
 Black hole entropy from an $SU(2)$-invariant formulation of Type~I isolated horizons,
\href{http://dx.doi.org/10.1103/PhysRevD.82.044050}{{\it Phys. Rev.~D}} {\bf 82} (2010), 044050, 23~pages,
\href{http://arxiv.org/abs/1006.0634}{arXiv:1006.0634}.

\bibitem{ash2}
Ashtekar A.,  Baez J.,  Corichi A.,  Krasnov K.,
Quantum geometry and black hole entropy,
\href{http://dx.doi.org/10.1103/PhysRevLett.80.904}{{\it Phys. Rev. Lett.}} {\bf 80} (1998), 904--907,
 \href{http://arxiv.org/abs/gr-qc/9710007}{gr-qc/9710007};\\
 Ashtekar A.,  Corichi A.,  Krasnov K.,
 Isolated horizons: the classical phase space,
 {\it Adv. Theor. Math. Phys.} {\bf 3} (2000), 419--478, \href{http://arxiv.org/abs/gr-qc/9905089}{gr-qc/9905089}; \\
 Ashtekar A.,   Baez J.,   Krasnov K.,
 Quantum geometry of isolated horizons and black hole entropy,
 {\it Adv. Theor. Math. Phys.} {\bf 4} (2000), 1--94, \href{http://arxiv.org/abs/gr-qc/0005126}{gr-qc/0005126}.


\bibitem{smo}
Smolin L.,
Linking topological quantum f\/ield theory and nonperturbative quantum gravity,
\href{http://dx.doi.org/10.1063/1.531251}{{\it J.~Math. Phys.}} {\bf 36} (1995), 6417--6455,
\href{http://arxiv.org/abs/gr-qc/9505028}{gr-qc/9505028}.

\bibitem{kras}
Krasnov K.V.,
On quantum statistical mechanics of Schwarzschild black hole,
\href{http://dx.doi.org/10.1023/A:1018820916342}{{\it  Gen. Relativity Gravitation}} {\bf 30} (1998), 53--68,
\href{http://arxiv.org/abs/gr-qc/9605047}{gr-qc/9605047}.

\bibitem{rov}
Rovelli C.,
Black hole entropy from loop quantum gravity,
\href{http://dx.doi.org/10.1103/PhysRevLett.77.3288}{{\it Phys. Rev. Lett.}} {\bf 77} (1996), 3288--3291,
 \mbox{\href{http://arxiv.org/abs/gr-qc/9603063}{gr-qc/9603063}}.

\bibitem{kmplb}
Kaul R.K., Majumdar P.,  	
Quantum black hole entropy,
\href{http://dx.doi.org/10.1016/S0370-2693(98)01030-2}{{\it Phys. Lett.~B}} {\bf 439} (1998), 267--270,
\href{http://arxiv.org/abs/gr-qc/9801080}{gr-qc/9801080}.

\bibitem{kmprl}
Kaul R.K., Majumdar P.,
Logarithmic correction to the Bekenstein--Hawking entropy,
\href{http://dx.doi.org/10.1103/PhysRevLett.84.5255}{{\it Phys. Rev. Lett.}} {\bf 84} (2000), 5255--5257,
\href{http://arxiv.org/abs/gr-qc/0002040}{gr-qc/0002040}.

\bibitem{dkmprd}
Das S., Kaul R.K.,   Majumdar P.,
A new holographic entropy bound from quantum geometry,
\href{http://dx.doi.org/10.1103/PhysRevD.63.044019}{{\it Phys. Rev.~D}} {\bf 63} (2001), 044019, 4~pages,
 \href{http://arxiv.org/abs/hep-th/0006211}{hep-th/0006211}.

\bibitem{krprd}
Kaul R.K., Kalyana Rama S.,
Black hole entropy from spin one punctures,
\href{http://dx.doi.org/10.1103/PhysRevD.68.024001}{{\it Phys. Rev.~D}} {\bf 68} (2003), 024001, 4~pages,
\href{http://arxiv.org/abs/gr-qc/0301128}{gr-qc/0301128}.

\bibitem{ms}
Fursaev D.V.,  	
Temperature and entropy of a quantum black hole and conformal anomaly,
\href{http://dx.doi.org/10.1103/PhysRevD.51.R5352}{{\it Phys. Rev.~D}} {\bf 51} (1995), R5352--R5355,
\href{http://arxiv.org/abs/hep-th/9412161}{hep-th/9412161}. \\
 Mann R.B., Solodukhin S.N.,
 Universality of quantum entropy for extreme black holes,
\href{http://dx.doi.org/10.1016/S0550-3213(98)00094-7}{{\it Nuclear Phys.~B}} {\bf523} (1998), 293--307,
\href{http://arxiv.org/abs/hep-th/9709064}{hep-th/9709064}.

\bibitem{witten}
Witten E.,  	
Quantum f\/ield theory and the Jones polynomial,
\href{http://dx.doi.org/10.1007/BF01217730}{{\it Comm. Math. Phys.}} {\bf 121} (1989), 351--399.

\bibitem{wbt}
Witten E., 	
On holomorphic factorization of WZW and coset models,
\href{http://dx.doi.org/10.1007/BF02099196}{{\it Comm. Math. Phys.}} {\bf 144} (1992), 189--212. \\
 Blau M., Thompson G.,  	
Derivation of the Verlinde formula from Chern--Simons theory and the G/G model,
\href{http://dx.doi.org/10.1016/0550-3213(93)90538-Z}{{\it Nuclear Phys.~B}} {\bf408} (1993), 345--390,
\href{http://arxiv.org/abs/hep-th/9305010}{hep-th/9305010}.

\bibitem{kaultop}
Kaul R.K.,
Chern--Simons theory, colored-oriented braids and link invariants,
\href{http://dx.doi.org/10.1007/BF02102019}{{\it Comm. Math. Phys.}} {\bf 162} (1994), 289--320,
\href{http://arxiv.org/abs/hep-th/9305032}{hep-th/9305032}. \\
 Kaul R.K.,
 Chern--Simons theory, knot invariants, vertex models and three manifold invariants,
in  Frontiers of Field Theory, Quantum Gravity and Strings,
{\it Horizons in World Physics}, Vol.~227, Nova Science Publi\-shers, New York, 1999, 45--63,
\href{http://arxiv.org/abs/hep-th/9804122}{hep-th/9804122}.\\
 Kaul R.K.,   Ramadevi P.,
 Three-manifold invariants from Chern--Simons f\/ield theory with arbitrary semi-simple gauge groups,
 \href{http://dx.doi.org/10.1007/s002200000347}{{\it Comm. Math. Phys.}} {\bf 217} (2001), 295--314,
 \href{http://arxiv.org/abs/hep-th/0005096}{hep-th/0005096}.

\bibitem{ver}
Di Francesco P.,   Mathieu P.,  Senechal D.,
 Conformal f\/ield theory, {\it Graduate Texts in Contemporary Physics},
Springer, Berlin, 1997.


\bibitem{liv1}
Livine E.R., Terno D.R., 	
 Quantum black holes: entropy and entanglement on the horizon,
\href{ttp://dx.doi.org/10.1016/j.nuclphysb.2006.02.012}{{\it Nuclear Phys.~B}} {\bf 741} (2006), 131--161,
 \href{http://arxiv.org/abs/gr-qc/0508085}{gr-qc/0508085}.

\bibitem{meiss}
Domagala M.,   Lewandowski J.,
Black-hole entropy from quantum geometry,
\href{http://dx.doi.org/10.1088/0264-9381/21/22/014}{{\it Classical Quantum Gravity}} {\bf 21} (2004), 5233--5243,
\href{http://arxiv.org/abs/gr-qc/0407051}{gr-qc/0407051}.  \\
  Meissner K.A.,
  Black-hole entropy in loop quantum gravity,
\href{http://dx.doi.org/10.1088/0264-9381/21/22/015}{{\it Classical Quantum Gravity}} {\bf 21} (2004), 5245--5251,
\href{http://arxiv.org/abs/gr-qc/0407052}{gr-qc/0407052}.  \\
 Khriplovich I.B.,
 Quantized black holes, correspondence principle, and holographic bound,
 \mbox{\href{http://arxiv.org/abs/gr-qc/0409031}{gr-qc/0409031}}.

\bibitem{gm}
Ghosh A., Mitra P.,
An improved lower bound on black hole entropy in the quantum geometry approach,
\href{http://dx.doi.org/10.1016/j.physletb.2005.05.003}{{\it Phys. Lett.~B}} {\bf 616} (2005), 114--117,
\href{http://arxiv.org/abs/gr-qc/0411035}{gr-qc/0411035}.

\bibitem{arefeva}
Aref'eva I.Ya.,
 Non-Abelian Stokes theorem,
\href{http://dx.doi.org/10.1007/BF01018469}{\textit{Theoret. and Math. Phys.}} \textbf{43} (1980), 353--356.

\bibitem{sahlmannthiemann}
Sahlmann H.,   Thiemann T.,
Chern--Simons theory, Stokes' theorem, and the Duf\/lo map,
\href{http://dx.doi.org/10.1016/j.geomphys.2011.02.013}{{\it J.~Geom. Phys.}} {\bf 61} (2011), 1104--1121,
\href{http://arxiv.org/abs/1101.1690}{arXiv:1101.1690}.\\
 Sahlmann H.,   Thiemann T.,
 Chern--Simons expectation values and quantum horizons from LQG and the Duf\/lo map,
 \href{http://arxiv.org/abs/1109.5793}{arXiv:1109.5793}.

\bibitem{duflo}
Duf\/lo M.,
Op\'erateurs dif\/fer\'entiels bi-invariants sur un groupe de Lie,
{\it Ann. Sci. \'Ecole Norm. Sup.~(4)}  {\bf 10} (1977), 265--288.


\bibitem{ca} Carlip S.,
Logarithmic corrections to black hole entropy from the Cardy formula,
\href{http://dx.doi.org/10.1088/0264-9381/17/20/302}{{\it Classical Quantum Gravity}} {\bf 17} (2000), 4175--4186,
\href{http://arxiv.org/abs/gr-qc/0005017}{gr-qc/0005017}.

\bibitem{cardy}
Cardy J.L.,
Operator content of two-dimensional conformally invariant theories,
\href{http://dx.doi.org/10.1016/0550-3213(86)90552-3}{{\it Nuclear Phys.~B}} {\bf270} (1986), 186--204, \\
 Bl\"ote H.W.J., Cardy J.L., Nightingale M.P.,
 Conformal invariance, the central charge, and universal f\/inite-size amplitudes at criticality,
\href{http://dx.doi.org/10.1103/PhysRevLett.56.742}{{\it Phys. Rev. Lett.}} {\bf 56} (1986), 742--745.

\bibitem{bhlms}
Brown J.D.,  Henneaux M.,
Central charges in the canonical realization of asymptotic symmetries: an example from three-dimensional gravity,
\href{http://dx.doi.org/10.1007/BF01211590}{{\it Comm. Math. Phys.}} {\bf 104} (1986), 207--226.\\
 Larsen F.,
 A string model of black hole microstates,
\href{http://dx.doi.org/10.1103/PhysRevD.56.1005}{{\it Phys. Rev.~D}} {\bf 56} (1997), 1005--1008,
 \href{http://arxiv.org/abs/hep-th/9702153}{hep-th/9702153}. \\
 Maldacena J.M.,   Strominger A.,
 Universal low-energy dynamics for rotating black holes,
  \href{http://dx.doi.org/10.1103/PhysRevD.56.4975}{{\it Phys. Rev.~D}} {\bf 56} (1997), 4975--4983,
  \href{http://arxiv.org/abs/hep-th/9702015}{hep-th/9702015}.

\bibitem{svh}
Strominger A.,  Vafa C.,
Microscopic origin of the Bekenstein--Hawking entropy,
 \href{http://dx.doi.org/10.1016/0370-2693(96)00345-0}{{\it Phys. Lett.~B}} {\bf 379} (1996), 99--104,
 \href{http://arxiv.org/abs/hep-th/9601029}{hep-th/9601029};\\
 Horowitz G.T,,  Lowe D.A., Maldacena J.M.,
 Statistical entropy of non extremal four-dimensional black holes and $U$ duality,
\href{http://dx.doi.org/10.1103/PhysRevLett.77.430}{{\it Phys. Rev. Lett.}} {\bf 77} (1996), 430--433,
\href{http://arxiv.org/abs/hep-th/9603195}{hep-th/9603195}.

 \bibitem{bs}
 Birmingham D., Sen S.,
 An exact black hole entropy bound,
\href{http://dx.doi.org/10.1103/PhysRevD.63.047501}{{\it Phys. Rev.~D}} {\bf 63} (2001), 047501, 3~pages,
\href{http://arxiv.org/abs/hep-th/0008051}{hep-th/0008051}. \\
Birmingham D.,  Sachs I.,  Sen S.,
Exact results for the BTZ black hole,
\href{http://dx.doi.org/10.1142/S0218271801001207}{{\it Internat. J. Modern Phys.~D}}  {\bf 10} (2001),  833--857,
\href{http://arxiv.org/abs/hep-th/0102155}{hep-th/0102155}.

 \bibitem{gks}
 Govindarajan T.R,,  Kaul R.K.,  Suneeta V.,  	
Logarithmic correction to the Bekenstein--Hawking entropy of the BTZ black hole,
\href{http://dx.doi.org/10.1088/0264-9381/18/15/303}{{\it Classical Quantum Gravity}} {\bf 18} (2001), 2877--2885,
\href{http://arxiv.org/abs/gr-qc/0104010}{gr-qc/0104010}.

 \bibitem{atwc}
Ach{\'u}carro A.,  Towensend P.K.,
 A Chern--Simons action for three-dimensional anti-de Sitter supergravity theories,
\href{http://dx.doi.org/10.1016/0370-2693(86)90140-1}{{\it Phys. Lett.~B}} {\bf 180} (1986), 89--92.\\
  Witten E.,
$(2+1)$-dimensional gravity as an exactly soluble system,
\href{http://dx.doi.org/10.1016/0550-3213(88)90143-5}{{\it Nuclear Phys.~B}} {\bf311} (1988), 46--78. \\
  Carlip S.,
Entropy from conformal f\/ield theory at Killing horizons,
\href{http://dx.doi.org/10.1088/0264-9381/16/10/322}{{\it Classical Quantum Gravity}} {\bf 16} (1999), 3327--3348,
\href{http://arxiv.org/abs/gr-qc/9906126}{gr-qc/9906126}.

 \bibitem{ct}
 Carlip S.,  Teitelboim C.,
 Aspects of black hole quantum mechanics and thermodynamics in $2+1$ dimensions,
\href{http://dx.doi.org/10.1103/PhysRevD.51.622}{{\it Phys. Rev.~D}} {\bf51} (1995), 622--631,
 \href{http://arxiv.org/abs/gr-qc/9405070}{gr-qc/9405070}.

 \bibitem{lab}
 Labastida J.M.F., Ramallo A.V.,
 Operator formalism for Chern--Simons theories,
\href{ttp://dx.doi.org/10.1016/0370-2693(89)91289-6}{{\it Phys. Lett.~B}} {\bf 227} (1989), 92--102. \\
 Isidro J.M., Labastida J.M.F., Ramallo A.V.,
 Polynomials for torus links from Chern--Simons gauge theories,
\href{http://dx.doi.org/10.1016/0550-3213(93)90632-Y}{{\it Nuclear Phys.~B}} {\bf398} (1993), 187--236,
 \href{http://arxiv.org/abs/hep-th/9210124}{hep-th/9210124}.\\
 Hayashi N.,
 Quantum Hilbert space of $G_C$ Chern--Simons--Witten theory and gravity,
\href{http://dx.doi.org/10.1143/PTPS.114.125}{{\it Prog. Theor. Phys. Suppl.}} (1993), no.~114, 125--147.

\bibitem{cskg} Carlip S.,
The statistical mechanics of the three-dimensional Euclidean black hole,
\href{http://dx.doi.org/10.1103/PhysRevD.55.878}{{\it Phys. Rev.~D}} {\bf 55} (1997), 878--882,
\href{http://arxiv.org/abs/gr-qc/9606043}{gr-qc/9606043}. \\
 Suneeta V.,  Kaul R.K.,  Govindarajan T.R.,
 BTZ black hole entropy from Ponzano--Regge gravity,
\href{http://dx.doi.org/10.1142/S0217732399000407}{{\it Modern Phys. Lett.~A}} {\bf 14} (1999), 349--358,
  \href{http://arxiv.org/abs/gr-qc/9811071}{gr-qc/9811071}.

 \bibitem{GKS}
 Govindarajan T.R.,   Kaul R.K.,  Suneeta V.,
 Quantum gravity on $dS_3$,
\href{http://dx.doi.org/10.1088/0264-9381/19/15/320}{{\it Classical Quantum Gravity}} {\bf 19} (2002), 4195--4205,
\href{http://arxiv.org/abs/hep-th/0203219}{hep-th/0203219}.

\bibitem{thooft}
't Hooft G.,
The black hole interpretation of string theory,
\href{http://dx.doi.org/10.1016/0550-3213(90)90174-C}{{\it Nuclear Phys.~B}} {\bf 335} (1990), 138--154.

\bibitem{suss}
Susskind L.,
Some speculations about black hole entropy in string theory,
\href{http://arxiv.org/abs/hep-th/9309145}{hep-th/9309145}.

\bibitem{hrs}
Halyo E.,  Rajaraman A.,  Susskind L.,  	
Braneless black holes,
\href{http://dx.doi.org/10.1016/S0370-2693(96)01544-4}{{\it Phys. Lett.~B}} {\bf 392} (1997), 319--322,
\href{http://arxiv.org/abs/hep-th/9605112}{hep-th/9605112}.\\
 Halyo E.,  Kol B.,  Rajaraman A.,   Susskind L.,
 Counting Schwarzschild and charged black holes,
 \href{http://dx.doi.org/10.1016/S0370-2693(97)00357-2}{{\it Phys. Lett.~B}} {\bf 401} (1997), 15--20,
   \href{http://arxiv.org/abs/hep-th/9609075}{hep-th/9609075}. \\
 Horowitz G.T.,   Polchinski J.,
 A correspondence principle for black holes and strings,
\href{http://dx.doi.org/10.1103/PhysRevD.55.6189}{{\it Phys. Rev.~D}} { \bf 55} (1997), 6189--6197,
\href{http://arxiv.org/abs/hep-th/9612146}{hep-th/9612146}.\\
 Damour T.,   Veneziano G.,
 Selfgravitating fundamental strings and black holes,
\href{http://dx.doi.org/10.1016/S0550-3213(99)00596-9}{{\it Nuclear Phys.~B}} {\bf 568} (2000), 93--119,
  \href{http://arxiv.org/abs/hep-th/9907030}{hep-th/9907030}. \\
 Halyo E.,
 Universal counting of black hole entropy by strings on the stretched horizon,
\href{http://dx.doi.org/10.1088/1126-6708/2001/12/005}{{\it J.~High Energy Phys.}} {\bf 2001} (2001), no.~12, 005, 15~pages,
 \href{http://arxiv.org/abs/hep-th/0108167}{hep-th/0108167}.

\bibitem{rkprd}
Kaul R.K.,  	
Black hole entropy from a highly excited elementary string,
\href{http://dx.doi.org/10.1103/PhysRevD.68.024026}{{\it Phys. Rev.~D}} {\bf 68} (2003), 024026, 4~pages,
\href{http://arxiv.org/abs/hep-th/0302170}{hep-th/0302170}.

 \bibitem{rama}
 Kalyana Rama S.,
Asymptotic density of open $p$-brane states with zero-modes included,
\href{http://dx.doi.org/10.1016/S0370-2693(03)00799-8}{{\it Phys. Lett.~B}} {\bf 566} (2003), 152--156,
\href{http://arxiv.org/abs/hep-th/0304152}{hep-th/0304152}.

 \bibitem{ggs}
 Gour G.,
 Algebraic approach to quantum black holes: logarithmic corrections to black hole entropy,
\href{http://dx.doi.org/10.1103/PhysRevD.66.104022}{{\it Phys. Rev.~D}} {\bf 66} (2002), 104022, 8~pages,
 \href{http://arxiv.org/abs/gr-qc/0210024}{gr-qc/0210024}.\\
  Gupta K.S., Sen S.,
  Further evidence for the conformal structure of a Schwarzschild black hole in an algebraic approach,
\href{http://dx.doi.org/10.1016/S0370-2693(01)01501-5}{{\it Phys. Lett.~B}} {\bf 526} (2002), 121--126,
  \href{http://arxiv.org/abs/hep-th/0112041}{hep-th/0112041}. \\
  Bianchi E.,
  Black hole entropy, loop gravity, and polymer physics,
\href{http://dx.doi.org/10.1088/0264-9381/28/11/114006}{{\it Classical Quantum Gravity}} {\bf 28} (2011), 114006, 12~pages,
\href{http://arxiv.org/abs/1011.5628}{arXiv:1011.5628}.

\bibitem{davidson}
Davidson A.,
Holographic shell model: stack data structure inside black holes,
\href{http://arxiv.org/abs/1108.2650}{arXiv:1108.2650}.


\bibitem{agui}
Agull{\'o} I., Barbero G. J.F., Borja E.F., Diaz-Polo J., Villase{\~n}or E.J.S.,
Combinatorics of the $SU(2)$ black hole entropy in loop quantum gravity,
\href{http://dx.doi.org/10.1103/PhysRevD.80.084006}{{\it Phys. Rev.~D}} {\bf 80} (2009), 084006, 3~pages,
  \href{http://arxiv.org/abs/0906.4529}{arXiv:0906.4529}. \\
Agull{\'o} I., Barbero G. J.F., Borja E.F., Diaz-Polo J., Villase{\~n}or E.J.S.,
 Detailed black hole state counting in loop quantum gravity,
 \href{http://dx.doi.org/10.1103/PhysRevD.82.084029}{{\it Phys. Rev.~D}} {\bf 82} (2010), 084029, 31~pages,
 \href{http://arxiv.org/abs/1101.3660}{arXiv:1101.3660}; \\
  Engle J.,  Noui K.,  Perez A., Pranzetti D.,
  The $SU(2)$ black hole entropy revisited,
 \href{http://dx.doi.org/10.1007/JHEP05(2011)016}{{\it J.~High Energy Phys.}}  {\bf 2011} (2011), no.~05, 016, 30~pages,
\href{http://arxiv.org/abs/1103.2723}{arXiv:1103.2723}.

\bibitem{liv2}
Freidel L.,  Livine E.R.,
The f\/ine structure of $SU(2)$ intertwiners from $U(N)$ representations,
 \href{http://dx.doi.org/10.1063/1.3473786}{{\it J.~Math. Phys.}}  {\bf 51} (2010), 082502, 19~pages,
 \href{http://arxiv.org/abs/0911.3553}{arXiv:0911.3553}.


\bibitem{sahl}
Sahlmann H.,
Black hole horizons from within loop quantum gravity,
\href{http://dx.doi.org/10.1103/PhysRevD.84.044049}{{\it Phys. Rev.~D}} {\bf 84} (2011), 044049, 12~pages,
\href{http://arxiv.org/abs/1104.4691}{arXiv:1104.4691}.

\bibitem{kr}
Krasnov K.,   Rovelli C.,
Black holes in full quantum gravity,
\href{http://dx.doi.org/10.1088/0264-9381/26/24/245009}{{\it Classical Quantum Gravity}} {\bf 26} (2009), 245009, 8~pages,
\href{http://arxiv.org/abs/0905.4916}{arXiv:0905.4916}.

\bibitem{abv}
Agull{\'o} I., Barbero G. J.F.,  D{\'{\i}}az-Polo J., Borja E.F., Villase{\~n}or E.J.S.,
 	Black hole state counting in LQG: a~number theoretical approach,
\href{http://dx.doi.org/10.1103/PhysRevLett.100.211301}{{\it Phys. Rev. Lett.}} {\bf 100} (2008), 211301, 4~pages,
\href{http://arxiv.org/abs/0802.4077}{arXiv:0802.4077}. \\
 Barbero G. J.F., Villase{\~n}or E.J.S.,
 Generating functions for black hole entropy in loop quantum gravity,
\href{http://dx.doi.org/10.1103/PhysRevD.77.121502}{{\it Phys. Rev.~D}} {\bf 77} (2008), 121502, 5~pages,
 \href{http://arxiv.org/abs/0804.4784}{arXiv:0804.4784}. \\
Barbero G.J.F., Villase{\~n}or E.J.S.,
On the computation of black hole entropy in loop quantum gravity,
\href{http://dx.doi.org/10.1088/0264-9381/26/3/035017}{{\it Classical Quantum Gravity}} {\bf 26} (2009), 035017, 22~pages,
\href{http://arxiv.org/abs/0810.1599}{arXiv:0810.1599}.


\bibitem{cm}
Caravelli F., Modesto L.,
Holographic actions from black hole entropy,
\href{http://dx.doi.org/10.1016/j.physletb.2011.07.023}{\textit{Phys. Lett.~B}} {\bf 702} (2011), 307--311,
\href{http://arxiv.org/abs/1001.4364}{arXiv:1001.4364}.


\bibitem{dmb}
Das S., Majumdar P., Bhaduri R.K.,
General logarithmic corrections to black hole entropy,
\href{http://dx.doi.org/10.1088/0264-9381/19/9/302}{{\it Classical Quantum Gravity}} {\bf 19} (2002), 2355--2367,
\href{http://arxiv.org/abs/hep-th/0111001}{hep-th/0111001}.

\bibitem{gourmed}
Gour G.,  Medved A.J.M.,
Thermal f\/luctuations and black hole entropy,
\href{http://dx.doi.org/10.1088/0264-9381/20/15/303}{{\it Classical Quantum Gravity}} {\bf 20} (2003), 3307--3326,
\href{http://arxiv.org/abs/gr-qc/0305018}{gr-qc/0305018}.

\bibitem{acpm}
 Chatterjee A.,  Majumdar P.,
 Universal canonical black hole entropy,
 \href{http://dx.doi.org/10.1103/PhysRevLett.92.141301}{{\it Phys. Rev. Lett.}} {\bf 92}  (2004), 141301, 4~pages,
 \href{http://arxiv.org/abs/gr-qc/0309026}{gr-qc/0309026}.  \\
 Chatterjee A.,  Majumdar P.,
 Universal criterion for black hole stability,
 \href{http://dx.doi.org/10.1103/PhysRevD.72.044005}{{\it Phys. Rev.~D}} {\bf 72} (2005), 044005, 3~pages,
  \href{http://arxiv.org/abs/gr-qc/0504064}{gr-qc/0504064}.\\
 Majumdar P.,
 Generalized Hawking--Page phase transition,
\href{http://dx.doi.org/10.1088/0264-9381/24/7/005}{{\it Classical Quantum Gravity}} {\bf 24} (2007), 1747--1753,
\href{http://arxiv.org/abs/gr-qc/0701014}{gr-qc/0701014}.

\bibitem{efs}
Easson D.A.,  Frampton P.H., Smoot G.F.,
 Entropic inf\/lation,
 \href{http://arxiv.org/abs/1003.1528}{arXiv:1003.1528}.

\end{thebibliography}
\end{document}